\documentclass[aps,prb,amsmath,amssymb,twocolumn,superscriptaddress]{revtex4-2}

\usepackage{graphicx}
\usepackage{dcolumn}
\usepackage{bm}

\setcitestyle{numbers,square}

\usepackage{graphicx}
\usepackage{dcolumn}
\usepackage{bm}
\usepackage{hyperref}

\usepackage[english]{babel}

\usepackage{tikz}
\usepackage{amssymb}

\begin{document}

\title{Dynamic dissipative structures in bistable magnetic ordered spin crossover systems: self-oscillations of magnetization}

\author{Yu.~S.~Orlov}
\affiliation{Siberian Federal University, 660041 Krasnoyarsk, Russia}
\affiliation{Kirensky Institute of Physics, Siberian Branch of the Russian Academy of Sciences, 660036 Krasnoyarsk, Russia}
\author{N.~N.~Paklin}
\affiliation{Siberian Federal University, 660041 Krasnoyarsk, Russia}

\author{S.~V.~Nikolaev}
\affiliation{Siberian Federal University, 660041 Krasnoyarsk, Russia}
\affiliation{Kirensky Institute of Physics, Siberian Branch of the Russian Academy of Sciences, 660036 Krasnoyarsk, Russia}
\author{E.~I.~Shneyder}
\affiliation{Kirensky Institute of Physics, Siberian Branch of the Russian Academy of Sciences, 660036 Krasnoyarsk, Russia}
\author{V.~A.~Dudnikov}
\affiliation{Kirensky Institute of Physics, Siberian Branch of the Russian Academy of Sciences, 660036 Krasnoyarsk, Russia}
\date{\today}

\begin{abstract}
Dissipative systems can exhibit a variety of behavioral modes, ranging from complex deterministic chaos to the spontaneous emergence of ordered structures. A simple example of the latter is Benard cells. More complex examples include lasers, droplet clusters, the Belousov--Zhabotinsky reaction, and biological life. Of particular interest in the context of the formation of spatiotemporal dissipative structures are bistable systems with spin crossover. This paper discusses the possibility of observing self-oscillations of magnetization in magnetically ordered systems with spin crossover. The results of theoretical calculations of the nonlinear dynamics of bistable magnetic systems under nonequilibrium conditions are presented.
\end{abstract}

\maketitle

\section{Introduction \label{intro}}
The phenomenon of spin transition, also known as spin crossover (SC), involves switching between the high-spin (HS) and low-spin (LS) states of a molecule and is observed in coordination compounds of transition metal ions with $d^4$–$d^7$ electron configurations. Spin crossover is accompanied by significant changes in the magnetic, electronic (optical), and structural properties of coordination compounds~\cite{Gutlich_book, SC_Cobaltite_book}, which underpins their high-tech potential in  information storage and processing devices, switches, sensors, indicators, and displays~\cite{Halcrow_book, ChemSocRev.40.3313.2011, Science.279.44.1998, AngewandteChemie.134.e202208208.2022, AdvancedMaterials.36.2307718.2024, CCSChemistry.5.915.2023, ChemicalPapers.77.7331.2023, ScientificReports.6.38334.2016, Magnetochemistry.10.107.2024, JMaterChemC.2.2949.2014, ChemistryMaterials.35.3276.2023, JournalAmericanChemicalSociety.144.14297.2022, SensorsActuatorsB.393.134147.2023, DaltonTrans.54.12432.2025, ChemistryMaterials.36.4889.2024, CrystEngComm.28.1707.2026}. More than 90 years have passed since the discovery~\cite{Cambi_1931} of the spin crossover effect by Cambi et al. During this time, hundreds of mono- and polynuclear complexes exhibiting spin crossover have been synthesized and studied in solid, liquid, and gel states. SC systems comprise a vast class of substances, including organic and inorganic, crystalline and non-crystalline (amorphous, polymeric, glassy) compounds, and continue to attract the attention of researchers from diverse fields: physics~\cite{NaturePhotonics.8.65.2018, NatureCommunications.15.6171.2024, APLMaterials.13.011105.2025, AppliedPhysicsLetters.124.122202.2024}, chemistry~\cite{NatureChemistry.14.739.2022, JournalSolidStateChemistry.304.122554.2021, ChemistryEuropeanJournal.30.e202403437.2024, ReviewsInorganicChemistry.46.35.2026}, biology~\cite{CoordinationChemistryReviews.344.363.2017, ChemistryEuropeanJournal.22.12176.2016, NatureCommunications.11.4145.2020, ACSOmega.3.9241.2018, RSCAdv.13.9020.2023, Science.356.1276.2017}, geophysics~\cite{NatureCommunications.12.2041.2021, Science.317.1740.2007, EarthPlanetaryScienceLetters.618.118296.2023, NatureCommunications.15.1961.2024}, and medicine~\cite{DaltonTrans.51.12427.2022, ScientificReports.8.14911.2018}. This is primarily due to the development and emergence of new experimental capabilities, such as the generation of ultra-strong magnetic fields and ultra-high pressures, the advancement of pump-probe spectroscopy with high temporal resolution, nanostructuring, and progress in electron-beam epitaxy and lithography, among others. The literature contains theoretical and experimental studies on nonlinear phenomena in open SC systems~\cite{EurophysLett.104.27003.2013, PhysRevB.89.224303.2014, J.Phys.Chem.C.122.20952.2018, ComptesRendusChimie.21.1133.2018, Crystals.9.46.2019, Magnetochem.5.21.2019, J.Phys.Chem.C.124.28093.2020, J.Applied.Phys.129.153901.2021, MaterialsTodayPhys.27.100842.2022, PhysRevB.105.174436.2022, JPhysChemSolids.190.111985.2024, PhysRevLett.110.087208.2013, Angew.Chem.Int.Ed.53.7539.2014, PhysRevB.89.024306.2014, Angew.Chem.Int.Ed.55.1755.2016, AdvancedTheorySimulations.1.1800080.2018, Opt.Express.21.31179.2013}. The theory of nonequilibrium states includes the concept of self-organization, a phenomenon in which a system becomes ordered far from equilibrium~\cite{Haken_book}. In some cases, self-organization leads to a state where Belousov--Zhabotinsky-type reactions occur in the system. Experimental studies of dissipative structures and observations of autocatalytic oscillations in SC systems can be found in the papers~\cite{AdvancedTheorySimulations.1.1800080.2018, Opt.Express.21.31179.2013}. The main innovation of SC systems is that the spontaneous formation of dissipative structures in them does not result from true diffusion of matter (as observed in Turing structures in hydrodynamics or biology) but rather from the effective diffusion of concentrations (populations) of the LS and HS multielectron states of the transition metal ion. Most studies of supercritical systems under nonequilibrium conditions focus on weakly magnetic coordination compounds. The formation of spatiotemporal dissipative structures in magnetically ordered systems (e.g., transition metal oxides~\cite{Lyubutin_UFN_2009}) near spin bistability has not yet been addressed in the literature and remains an open question. This paper presents theoretical modeling of the formation mechanisms of such spatiotemporal structures arising under highly nonequilibrium conditions. These results reveal a form of supramolecular coherent behavior of a large number of transition metal ions, giving rise to temporal and spatial oscillations of both the magnetization of the substance and the populations of electronic terms with different multiplicities. Nonlinear phenomena of the Belousov--Zhabotinsky reaction type (autocatalytic oscillations of magnetization) are considered in open systems with spin crossover near bistability.

\section{Effective Hamiltonian \label{Model}}
Let us consider a crystal lattice in which transition metal cations are located at each site in the crystal field of ligands. Instead of the full set of multielectron terms of $3d$~-~ions, we will restrict ourselves to only two low-energy terms, between which crossover is possible with increasing crystal field $\Delta = 10Dq$. For definiteness, we will focus on transition metal ions with a $3d^6$ electron configuration. According to the Tanabe-Sugano diagrams~\cite{TanabeSugano_JPSJ.9.766.1954}, in this case the ground electron term of the cation, i.e., the term with minimum energy, can be either the HS term ${}^5{T_{2g}}$ with spin $S = 2$, or the LS term ${}^1{A_{1g}}$ with spin $S = 0$. To describe cooperative phenomena in SC systems, it is convenient to use the representation of the pseudospin vector operator $\hat \tau _i = \left( {{{\hat \tau }^x_i},{{\hat \tau }^y_i},{{\hat \tau }^z_i}} \right)$, defined at each lattice site $i$ in the orbital subspace of the Hilbert space of many-electron HS and LS quantum-mechanical states. The pseudospin projection operator reads
\begin{equation}
\label{sigma}
{\hat \tau ^z _i} = \frac{1}{2}\left( {{X^{HS,HS}_i} - {X^{LS,LS}_i}} \right), \nonumber
\end{equation}
where ${X^{HS,HS}_i} = \left| {HS} \right\rangle \left\langle {HS} \right|$ and ${X^{LS,LS}_i} = \left| {LS} \right\rangle \left\langle {LS} \right|$ are Hubbard $X$-operators satisfying the completeness condition ${X^{LS,LS}_i} + {X^{HS,HS}_i} = 1$. The operator~\ref{sigma} has two eigenvalues $\sigma = \pm \frac{1}{2}$ corresponding to the HS and LS states: ${\hat \tau ^z _i}\left| {HS} \right\rangle  = \frac{1}{2}\left| {HS} \right\rangle $, ${\hat \tau ^z _i}\left| {LS} \right\rangle  =  - \frac{1}{2}\left| {LS} \right\rangle $. Using the operator ${\hat \tau ^z _i}$ defined in this way, the occupation number operators can be written as ${\hat n^{LS}_{i}} = - {\hat \tau ^z _i} + \frac{1}{2}$ and ${\hat n^{HS}_i} = {\hat \tau ^z _i} + \frac{1}{2}$ so that ${\hat n^{LS}_i} + {\hat n^{HS}_i} = 1$ holds by construction. The components ${\hat \tau ^x _i}$ and ${\hat \tau ^y _i}$ are given by
\begin{equation}
{\hat \tau ^x _i} =   \frac{1}{2}\left( {{{\hat \tau }^ + _i} + {{\hat \tau }^ - _i}} \right), \\ \nonumber
{\hat \tau ^y _i} = - \frac{1}{2}\left( {{{\hat \tau }^ + _i} - {{\hat \tau }^ - _i}} \right), \\ \nonumber
\end{equation}
where ${\hat \tau ^ + _i} = {X^{HS,LS}_i} = \left| {HS} \right\rangle \left\langle {LS} \right|$ and ${\hat \tau ^ - _i} = {X^{LS,HS}_i} = \left| {LS} \right\rangle \left\langle {HS} \right|$ are the pseudospin projection raising and lowering operators: ${\hat \tau ^ + _i}\left| {LS} \right\rangle  = \left| {HS} \right\rangle $, ${\hat \tau ^ - _i}\left| {HS} \right\rangle  = \left| {LS} \right\rangle $. It is easy to check that the $\tau$-operators satisfy the same commutation relations as the usual spin operators:
\begin{equation}
\left[ {{{\hat \tau }^ + _i},{{\hat \tau }^ - _i}} \right] = 2{\hat \tau ^z _i}, \\ \nonumber
\left[ {{{\hat \tau }^z _i},{{\hat \tau }^ \pm _i}} \right] =  \pm {\hat \tau ^ \pm _i}. \\ \nonumber
\end{equation}
The Hamiltonian of the two-level model for non-interacting cation-anion complexes, or SC complexes, on a lattice can be written in the $\tau$-representation as
\begin{equation}
    \hat H_\tau = \Delta_{\tau}\sum\limits_i {\hat \tau _i^z}.
    \label{H_tau}
\end{equation}
Here ${\Delta _\tau } = {F_{HS}} - {F_{LS}}$ is the difference between the free energies of the ion in the HS and LS states. ${F_{HS\left( {LS} \right)}} = {E_{HS\left( {LS} \right)}} - {k_B}T{S_{HS\left( {LS} \right)}}$, where ${E_{HS\left( {LS} \right)}}$ is the energy of the corresponding state. $E_{HS} = -4Dq - 21B$, $E_{LS} \approx -24Dq - 16B + 8C$, with $B$ and $C$ being the Racah parameters ($B = \gamma C$). The characteristic values of the Racah parameters for $3d^6$ ions can be found in~\cite{TanabeSugano_JPSJ.9.766.1954}: $B = 1065~\text{cm}^{-1} = 0.132~\text{eV}$, $\gamma = 4.8$. $S_{HS\left( LS \right)} = \ln {g_{HS\left( LS \right)}}$ is the entropy of the ion in the HS (LS) state and $g_{HS\left( LS \right)}$ is the degeneracy factor of the HS(LS) terms. ${\Delta _\tau } = {\Delta _S} - {k_B}T\ln g$, where ${\Delta _S} = E_{HS} - E_{LS} = 2\left(\Delta - \Delta_0\right)$ is the spin gap ($2\Delta_0 = 5B + 8C$), $g = \frac{g_{HS}}{g_{LS}}$ is the ratio of degeneracy multiplicities. At the crossover point, $\Delta_S = 0$ when $\Delta = \Delta_0$. The degeneracy multiplicity of each of the two spin states can be written as the product ${g_{HS\left( {LS} \right)}} = {g_{\tau ,HS\left( {LS} \right)}} \cdot {g_{S,HS\left( {LS} \right)}}$, where ${g_{\tau ,HS\left( {LS} \right)}}$ and ${g_{S,HS\left( {LS} \right)}}$ are the orbital and spin degeneracy multiplicities of the HS(LS) state. Therefore, $g = {g_\tau } \cdot {g_S}$, where ${g_\tau } = \frac{{{g_{\tau ,HS}}}}{{{g_{\tau ,LS}}}}$ and ${g_S} = \frac{{{g_{S,HS}}}}{{{g_{S,LS}}}}$ is the ratio of orbital and spin degeneracy multiplicities. In the case under consideration, for the LS and HS states of the $3d^6$ ion, ${g_{S,LS}} = 1$ and ${g_{S,HS}} = 5$, respectively. The orbital degeneracy multiplicity ${g_{\tau ,HS\left( {LS} \right)}}$ is generally determined not only by the dimension of the irreducible representation used to classify the multielectron terms ${}^5{T_{2g}}$ and ${}^1{A_{1g}}$, but also by the number of vibronic states.

In spin crossover compounds, the interplay between magnetic, thermal, and electronic (optical) properties is clearly manifested. Therefore, in most cases, their description requires taking into account the interaction between different degrees of freedom (subsystems). By including the electron-phonon interaction, we obtain:
\begin{equation}
    \hat H = \hat H_\tau + \hat H_{\tau  - ph},
    \label{H}
\end{equation}
where
\begin{eqnarray}
    &{\hat H_{\tau  - ph}} = \sum\limits_i {\left( {\frac{{\hat p_i^2}}{{2M}} +
    \frac{1}{2}{k_0}\hat u_i^2} \right)}  - \frac{1}{2}{V_u}\sum\limits_{\left\langle {i,j} \right\rangle } {{{\hat u}_i}{{\hat u}_j}} \nonumber  \\
    &- \sum\limits_i {\left( {{g_1}{{\hat u}_i} + {g_2}\hat u_i^2} \right)\hat \tau _i^z}.
    \label{H_tau-ph}
\end{eqnarray}
Here, the first term is the energy of the local fully symmetric vibrations of the cation-anion complex. Among all the vibrational modes of the SC complex, we consider only the breathing $A_{1g}$ mode, since it is most strongly coupled to the spin multiplicity fluctuations of the $3d$ ions. The $A_{1g}$ mode of the ligands can be modeled as a harmonic oscillator with mass $M$ and elastic constant $k_0$. The operator $\hat u$ is the displacement operator of the harmonic oscillator, or of the ligands, from the equilibrium position. The operator $\hat p$ is the momentum operator conjugate to $\hat u$. The second and third terms in (\ref{H_tau-ph}) describe, respectively, the elastic interaction of cations at neighboring lattice sites and the electron-vibrational (vibronic) interaction. Here, ${V_u}$ is the parameter of elastic intermolecular interaction, and ${g_1}$ and  ${g_2}$ are the electron-vibrational coupling constants.

The contribution to the electron-phonon interaction in Eq.~(\ref{H_tau-ph}) that is linear in the displacement operator $\hat u$ leads to a difference in the metal-ligand bond lengths in the LS and HS states. Indeed, the different signs in front of the operators ${X^{HS,HS}_i}$ and ${X^{LS,LS}}_i$  in the electron-vibronic interaction correspond to the opposite influence of the displacement $u = \left\langle {\hat u} \right\rangle $ on the energy of these states. Here and below, the angular brackets  $\left\langle \ldots \right\rangle $  denote the quantum-statistical average. An increase (decrease) in $u$ leads to a decrease (increase) in the crystal field and stabilization of the HS (LS) state. Therefore, filling of the LS state is associated with a decrease in the cation-anion bond length, while filling of the HS state is conversely associated with an increase. The metal-ligand bond length can be represented as $l\left( T \right) = {l_0}\left( T \right) + u\left( T \right)$, where ${l_0}\left( T \right)$ is the regular component due to the anharmonicity of lattice vibrations, and the anomalous contribution  $u\left( T \right)$ arises due to the vibronic interaction ${g_1}$. It is easy to show that at $T = 0$ in the absence of LS-HS spin-orbit interaction, ${l_{LS\left(HS\right)}} = {l_0} + u_{LS\left(HS\right)}^0$, where $u_{LS}^0 = {{ - {g_1}} \mathord{\left/ {\vphantom {{ - {g_1}} {{k_0}}}} \right. \kern-\nulldelimiterspace} {{k_0}}}$ and $u_{HS}^0 = {{{g_1}} \mathord{\left/ {\vphantom {{{g_1}} {{k_0}}}} \right. \kern-\nulldelimiterspace} {{k_0}}}$.
Using the estimated values $g_1 = 0.8$~eV/\AA\ and ${k_0} = 7.5$~eV/\AA$^2$~\cite{Phys.Rev.B.84.104119.2011}, we obtain $u_{LS}^0 =  - 0.09$~\AA, $u_{HS}^0 = 0.13$~\AA\ and thus $\Delta {u_0} = u_{HS}^0 - u_{LS}^0 = 0.22$~\AA. Since the bond length ${l_0}$ at $T = 0$ is about 2~\AA, $\Delta {u_0}$ is 10\% of this value. It is seen that in the absence of electron-vibrational interaction, $u_{LS}^0 = u_{HS}^0= 0$, therefore, a change in the system's volume with increasing temperature is possible only due to lattice anharmonicity. It is known that the thermal expansion coefficient of rare earth cobalt oxides exhibits an unusual temperature dependence: two anomalies associated with the population of the HS state at low temperatures and the dielectric-metal transition at high temperatures~\cite{Eur.Phys.J.B.47.213.2005,Phys.Rev.B.78.134402.2008,Phys.Rev.B.66.094408.2002,Phys.Rev.B.71.014443.2005}. These anomalies are most pronounced in LaCoO$_3$. From the foregoing, it can be inferred that the linear electron-phonon interaction $g_1$ is responsible for the low-temperature anomaly.

Using decoupling $\hat u_i^2X_i^{LS,LS} \approx \hat u_i^2\left\langle {X_i^{LS,LS}} \right\rangle $ and $\hat u_i^2X_i^{HS,HS} \approx \hat u_i^2\left\langle {X_i^{HS,HS}} \right\rangle $, from (\ref{H_tau-ph}) we get
\begin{eqnarray}
{\hat H_{\tau - ph}} = \sum\limits_i {\left( {\frac{{\hat p_i^2}}{{2M}} + \frac{1}{2}\left[ {{k_0} + 2{g_2}\left( {1 - 2{n_{HS}}} \right)} \right]\hat u_i^2} \right)} \nonumber \\
- \frac{1}{2}{V_u}\sum\limits_{\left\langle {i,j} \right\rangle } {{{\hat u}_i}{{\hat u}_j}} 
- {g_1}\sum\limits_i {{{\hat u}_i}{\hat \tau _i^z}},
\label{H_tau-ph_u2decoupling}
\end{eqnarray}
where ${n_{LS}} = \left\langle {X_i^{LS,LS}} \right\rangle $ and ${n_{HS}} = \left\langle {X_i^{HS,HS}} \right\rangle $ are the populations of the LS and HS states. From Eq.~(\ref{H_tau-ph_u2decoupling}) it is clear that the frequencies of local oscillations are ${\omega _{HS\left( {LS} \right)}} = \sqrt {{{{k_{HS\left( {LS} \right)}}} \mathord{\left/  {\vphantom {{{k_{HS\left( {LS} \right)}}} M}} \right. \kern-\nulldelimiterspace} M}} $, where ${k_{HS}} = {k_0} - 2{g_2}$ and ${k_{LS}} = {k_0} + 2{g_2}$. Thus, the frequencies in the HS (${n_{HS}} = 1$) and LS (${n_{HS}} = 0$) states are different, with ${\omega _{HS}} < {\omega _{LS}}$.

Using the Fourier transform  ${\hat u_f} = \frac{1}{{\sqrt N }}\sum\limits_{\bm q} {{{\hat A}_{\bm q}}{e^{i\bm qR_f}}} $, ${\hat p_f} = \frac{1}{{\sqrt N }}\sum\limits_{\bm q} {{{\hat P}_{\bm q}}{e^{ - i{\bm q}R_f}}} $, where $N$ is the number of lattice sites, and the canonical transformation 
$$
{\hat A_{\bm q}} = \sqrt {\frac{ 1 }{2M\omega_{\bm q}}} \left( {{b_{\bm q}} + b_{ - \bm q}^\dag } \right),
$$
$$
{\hat P_{\bm q}} = i\sqrt {\frac{M \omega_{\bm q}}{2}} \left( {{b_{\bm q}} - b_{ - \bm q}^\dag } \right),
$$
the Hamiltonian \eqref{H_tau-ph_u2decoupling} can be reduced to the form
\begin{eqnarray}
    {\hat H_{\tau  - ph}} = \sum\limits_{\bm q} {{\omega_{\bm q}}\left( {b_{\bm q}^\dag {b_{\bm q}} + \frac{1}{2}} \right)} \nonumber \\
    - \frac{1}{{\sqrt N }}\sum\limits_{i,\bm q} {\left( {{g_{i\bm q}}{b_{\bm q}} + g_{i\bm q}^ * b_{\bm q}^\dag } \right)\hat \tau _i^z},
    \label{H_tau-ph_inb}
\end{eqnarray}
where ${g_{i\bm q}} = {g_{\bm q}}{e^{i\bm q \cdot {R_i}}}$, ${g_{\bm q}} = \frac{1}{\sqrt {2M\omega_{\bm q}}}{g_1}$. 
For a simple cubic lattice along the $\left[ {1,1,1} \right]$ direction, the phonon dispersion is given by
$$
{\omega_{\bm q}} = \sqrt {{\omega ^2}\left[ {1 - \frac{{{V_u}}}{{3{k_0}}}\left( {\cos {q_x} + \cos {q_y} + \cos {q_z}} \right)} \right]}.
$$
Here, $\omega  = \sqrt {\frac{k}{M}} $, with $k = {k_{LS}} - 4{g_2}{n_{HS}}$, or
\begin{equation}
\omega  = \sqrt {\omega _{LS}^2 - {n_{HS}}\frac{{4{g_2}}}{M}}  = \sqrt {\omega _{LS}^2 - {n_{HS}}\Delta {\omega ^2}},
\label{Omega}
\end{equation}
where $\Delta {\omega ^2} = \omega _{LS}^2 - \omega _{HS}^2$.
From (\ref{Omega}) it is evident that the contribution to the electron-phonon interaction in (\ref{H_tau-ph}), quadratic in the displacement $\hat u$, describes the softening of the phonon spectrum. For the topic discussed in this paper, this effect is of no relevance, so we set $g_2=0$ henceforth. The softening of the phonon spectrum in rare-earth cobalt oxides caused by fluctuations in the multiplicity of Co$^{3+}$ ions is discussed in more detail in~\cite{JETPLett.115.615.2022}.

Using the Lang–Firsov unitary transformation ${\hat H^{eff}} = {e^{\hat U}}\hat H{e^{ - \hat U}}$, with
$$
\hat U = \frac{1}{{\sqrt N }}\sum\limits_{i,\bm q} {\omega_{\bm q}^{ - 1}\left( {{g_{i\bm q}}{b_{\bm q}} - g_{i\bm q}^ * b_{\bm q}^\dag } \right)\hat \tau _i^z}, 
$$
the Hamiltonian \eqref{H} is transformed into the effective Hamiltonian
\begin{eqnarray}
    {\hat H^{eff}} = {\Delta_\tau }\sum\limits_i {\hat \tau _i^z}  - \frac{1}{2}\sum\limits_{i,j} {{J_\tau }\left( {i,j} \right)\hat \tau _i^z\hat \tau _j^z},
    \label{H_eff}
\end{eqnarray}
which acts in the $\tau $-subspace of the Hilbert space and describes the elastic interaction
${J_\tau }\left( {i,j} \right) = \frac{1}{N} \sum\limits_{\bm q} {\frac{{{g_{i\bm q}}g_{j\bm q}^ * }}{{{\omega_{\bm q}}}}} =\frac{1}{N} \sum\limits_{\bm q} {\frac{{g_{\bm q}^2}}{{{\omega_{\bm q}}}}{e^{i\bm q \cdot \left( {{R_i} - {R_j}} \right)}}} 
$ between $3d$ ions at different sites of the crystal lattice. We will take into account only nearest-neighbor interactions.

The electron-phonon interaction in \eqref{H} can be eliminated not only by the Lang–Firsov transformation. To this end, it is convenient to pass to new operators~\cite{Gehring_ReportsProgressPhysics.38.1.2001} 
\begin{eqnarray}
\begin{split}
&\gamma_{\bm q}^\dag = b_{\bm q}^\dag - \frac{g_{\bm q}}{\omega_{\bm q}}\hat \tau _{-\bm q}^z, \\
&{\gamma _{\bm q}} = {b_{\bm q}} - \frac{g_{\bm q}}{\omega_{\bm q}}\hat \tau _{\bm q}^z,
\end{split}
\label{gamma_new}
\end{eqnarray}
which satisfy the same commutation relations as the original phonon creation and annihilation operators $b_{\bm q}^\dag $ and ${b_{\bm q}}$:
$$
\left[ {{\gamma_{\bm q'}},\gamma_{\bm q}^\dag } \right] = \left[ {{b_{\bm q'}},b_{\bm q}^\dag } \right] = {\delta_{\bm q\bm q'}}.
$$
Then, substituting \eqref{gamma_new} into \eqref{H_tau-ph_inb} yields
\begin{eqnarray}
    {\hat H_{\tau - ph}} = \sum\limits_{\bm q} {\omega_{\bm q}\left( {\gamma_{\bm q}^\dag {\gamma_{\bm q}} + \frac{1}{2}} \right)} - \sum\limits_{\bm q} {\frac{{g_{\bm q}^2}}{{{\omega_{\bm q}}}}\hat \tau _{\bm q}^z\hat \tau_{ - \bm q}^z}.
\end{eqnarray}
Using \eqref{gamma_new},
\begin{eqnarray}
    \left\langle {\gamma_{ - \bm q}^\dag  + \gamma_{\bm q}} \right\rangle = \left\langle {b_{ - \bm q}^\dag  + b_{\bm q}} \right\rangle - \frac{2g_{\bm q}}{\omega_{\bm q}}\left\langle {\hat \tau _{\bm q}^z} \right\rangle.
\label{gamma_b_tau}
\end{eqnarray}
Since $\left\langle {\gamma_{ - \bm q}^\dag  + \gamma_{\bm q}} \right\rangle = 0$, it follows from \eqref{gamma_b_tau} that $\left\langle {b_{ - \bm q}^\dag  + b_{\bm q}} \right\rangle = \frac{2g_{\bm q}}{\omega_{\bm q}}\left\langle {\hat \tau _{\bm q}^z} \right\rangle$.

For magnetically ordered SC systems, most of which are antiferromagnets, in addition to the interatomic elastic interaction ${J_\tau }$, the exchange interaction ${J_S}$ must be taken into account. Instead of \eqref{H_eff}, in this case one should consider the Hamiltonian
\begin{eqnarray}
    {\hat H^{eff}} = {\Delta _\tau }\sum\limits_i {\hat \tau _i^z}  - \frac{1}{2}{J_\tau }\sum\limits_{\left\langle {i,j} \right\rangle } {\hat \tau _i^z\hat \tau _j^z} \nonumber \\
    + \frac{1}{2}{J_S}\sum\limits_{\left\langle {i,j} \right\rangle } {{{\hat n}_{HS,i}}{{\hat n}_{HS,j}}{\hat {\bm S}_i} \cdot {\hat {\bm S}}_j},
    \label{H_eff_SS}
\end{eqnarray}
where the last term describes the exchange interaction between transition metal ions in the HS state and is analogous to the Heisenberg Hamiltonian for binary alloys. It should be noted that, in contrast to \eqref{H_eff}, the spin degrees of freedom in \eqref{H_eff_SS} are taken into account explicitly, so ${\Delta _\tau } = {\Delta _S} - {k_B}T\ln g_\tau$, if $J_S \ne 0$.

\section{Mean-field approximation \label{MFA}}
Let us introduce the vector operator ${\hat {\bm \Omega} _i} = \left( {\begin{array}{*{20}{c}}
{{{\hat n}_{HS,i}}\hat S_i^z}\\ 
{\hat \tau _i^z}
\end{array}} \right)$. Then, in the mean-field (MF) approximation for two sublattices $A$ and $B$ normalized to the number of magnetic cells  $N_C=N/2$, the Hamiltonian \eqref{H_eff_SS} takes the form
\begin{eqnarray}
    \hat H_{MF}^{eff} = \sum\limits_C {\hat H_C^{ion}} + {H_0},
    \label{H_eff_MF}
\end{eqnarray}
where $C = A, B$ is the sublattice index ($\bar C = B$, if $C = A$, and vice versa).
\begin{eqnarray}
    \hat H_C^{ion} = {\bm D_{\bar C}} \cdot {\hat {\bm \Omega} _C}
    \label{Hion_C}
\end{eqnarray}
is the energy of an ion in the resulting mean field ${\bm D_C} = \left( {{D_{S,C}},{D_{\tau ,C}}} \right)$. Here ${D_{S,C}} = z{J_S}{n_{HS,C}}{m_C}$, ${D_{\tau , C}} = {\Delta _\tau } - z{J_\tau }{\tau _C}$, where $z$ is the number of nearest neighbors, ${m_C} = \left\langle {\hat S_{{i_C}}^z} \right\rangle $ is the sublattice magnetization, and ${\tau _C} = \left\langle {\hat \tau _{{i_C}}^z} \right\rangle $ (${n_{HS,C}} = {\tau _C} + \frac{1}{2}$). The last term in \eqref{H_eff_MF} is
\begin{eqnarray}
    {H_0} =  - z{J_S}{n_{HS,A}}{n_{HS,B}}{m_A}{m_B} - z{J_\tau }{\tau _A}{\tau _B}.
\end{eqnarray}
In the MF approximation, it is convenient to redefine $zJ_S = J_S$ and $zJ_\tau = J_\tau$.

By solving the self-consistent eigenvalue problem for two sublattices
\begin{eqnarray}
    \hat H_C^{ion}\left| {{\varphi _{k,C}}} \right\rangle  = {E_k}\left| {{\varphi _{k,C}}} \right\rangle,
    \label{eig}
\end{eqnarray}
one can find the quantities of interest, namely the averages $m_C$ and ${\tau _C}$, entering the field ${\bm D_C}$:
$$
{m_C} = \frac{1}{Z_C}\sum\limits_k {\left\langle {{\varphi _{k,C}}\left| {\hat S_C^z} \right|{\varphi _{k,C}}} \right\rangle {e^{ - {E_k}\beta }}},
$$
$$
{\tau _C} = \frac{1}{Z_C}\sum\limits_k {\left\langle {{\varphi _{k,C}}\left| {\hat \tau _C^z} \right|{\varphi _{k,C}}} \right\rangle {e^{ - {E_k}\beta }}}. 
$$
Here $\beta  = {1 \mathord{\left/
 {\vphantom {1 {{k_B}T}}} \right.
 \kern-\nulldelimiterspace} {{k_B}T}}$,
$$
Z_C = \sum\limits_k {{e^{ - {E_k}\beta }}} = {e^{ - \beta \frac{1}{2}{D_{\tau ,C}}}}\left( {{e^{\beta {D_{\tau ,C}}}} + {\theta_C}} \right) 
$$
is the partition function for sublattice $C$, where
$$
{\theta_C} = \frac{{\sinh \left[ {{{\left( {2S + 1} \right)\beta {D_{S,C}}} \mathord{\left/
{\vphantom {{\left( {2S + 1} \right)\beta {D_{S,C}}} 2}} \right.
\kern-\nulldelimiterspace} 2}} \right]}}{{\sinh \left( {{{\beta {D_{S,C}}} \mathord{\left/
{\vphantom {{\beta {D_{S,C}}} 2}} \right.
\kern-\nulldelimiterspace} 2}} \right)}}.
$$
In the calculation process, we select the solutions that correspond to the local and global minima of the free energy $F = - {\beta ^{ - 1}}\ln Z + H_0$, where $Z = Z_A \cdot Z_B$. The eigenfunctions $\left| {{\psi _k}} \right\rangle $ of the Hamiltonian \eqref{H_eff_MF} can be written as a direct product of the states $\left| {{\varphi _{k,C}}} \right\rangle $ for different sublattices: $\left| {{\psi _k}} \right\rangle  = \left| {{\varphi _{k,A}}} \right\rangle \left| {{\varphi _{k,B}}} \right\rangle $. The states $\left| {{\varphi _{k,C}}} \right\rangle $ can in turn be represented as a linear combination 
$$
\left| {{\varphi _{k,C}}} \right\rangle  = {C_{LS}}{\left| {LS} \right\rangle _C} + \sum\limits_{s = - S}^{ + S} {C_{HS}^\sigma {{\left| {HS,s} \right\rangle }_C}}, 
$$
where $\left| {HS,s} \right\rangle  = \left| {HS} \right\rangle \left| s  \right\rangle $. The basis states $\left| s \right\rangle $ are eigenstates of the spin projection operator ${\hat S^z}$: ${\hat S^z}\left| s \right\rangle  = s \left| s \right\rangle $, with $s = - S, - S + 1,.. + S$.

Numerical calculations for the case $J_\tau > 0$ and $J_S > 0$, lead to the following results: ${\tau _C} = {\tau _{\bar C}} = \tau$ (or ${n_{HS,C}} = {n_{HS,\bar C}} = n_{HS}$) and ${m_C} =  - {m_{\bar C}}$.

\begin{figure}
\centering
\includegraphics[width=9.0cm]{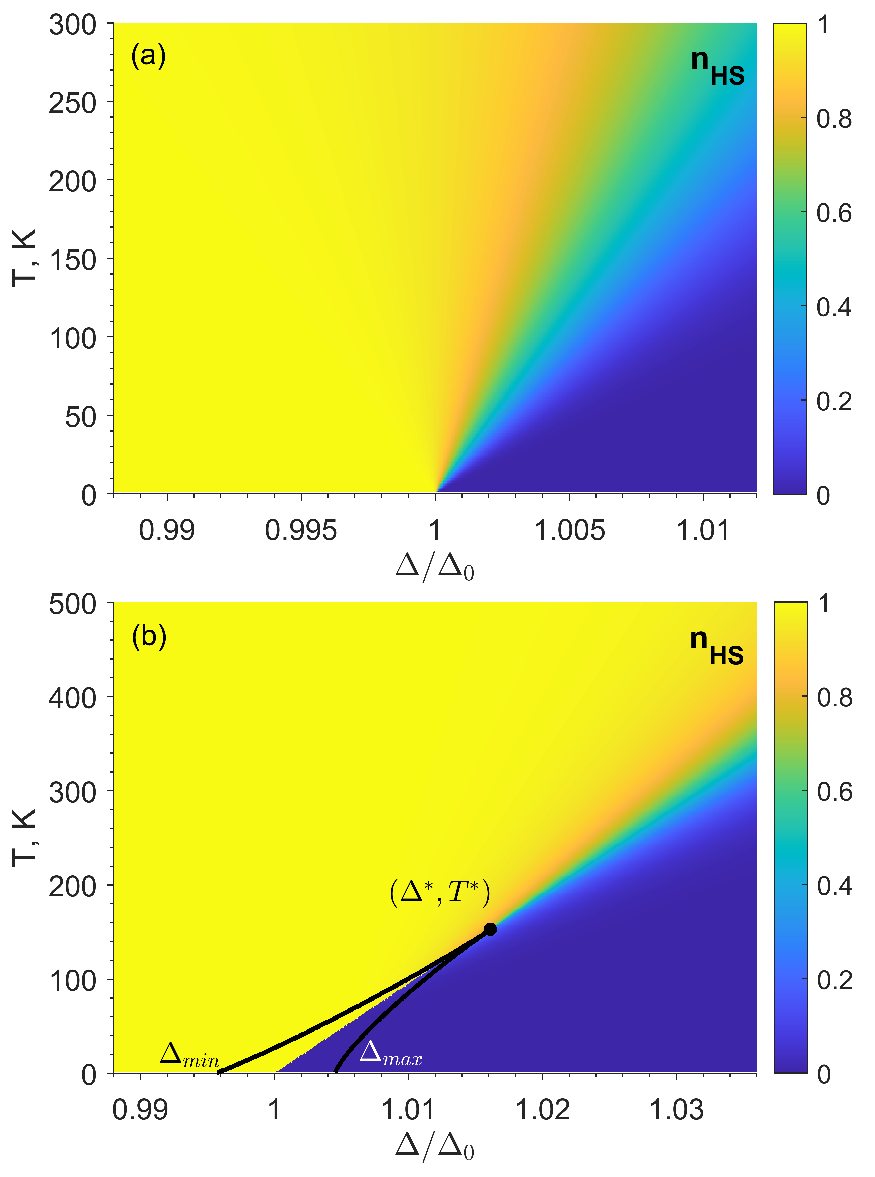}
\caption{\label{fig1} ${\Delta} - T$ phase diagrams of the HS state population ${n_{HS}}$ for ${J_S} = {J_\tau } = 0$~(a) and for ${J_S} = 0$, ${J_\tau } = 152$~K~(b). $\left( {\Delta^ * ,{T^*}} \right)$ is the critical point. ${\Delta}\min /\max $ are the boundaries of the metastable region at $T = 0$. The calculations were performed with $g = 15$.}
\end{figure}

To understand the mechanisms of cooperativity in SC systems, we consider each of the interactions $J_\tau$ and $J_S$ separately. Figure~\ref{fig1} shows calculated phase diagrams of the HS state population ${n_{HS}}$ in the temperature $T$ versus crystal field ${\Delta}$ coordinates in the absence of all interatomic interactions (Fig.~\ref{fig1}a) and in the presence of only elastic interatomic interaction ${J_\tau }$ (Fig.~\ref{fig1}b). It can be seen that the elastic interaction transforms the smooth spin crossover (Fig.~\ref{fig1}a) into a first-order phase transition at low temperatures (Fig.~\ref{fig1}b). The black solid lines in Fig.~\ref{fig1}b indicate the boundaries of the region of metastable states. The first-order phase transition line ends at the critical point $\left( {\Delta^* ,{T^*}} \right)$ (Fig.~\ref{fig1}b). As in the case of a liquid and gas, the SC system can be continuously converted from the HS to LS state and vice versa by going around the point $\left( {\Delta^* ,{T^*}} \right)$.

\begin{figure}
\centering
\includegraphics[width=9.0cm]{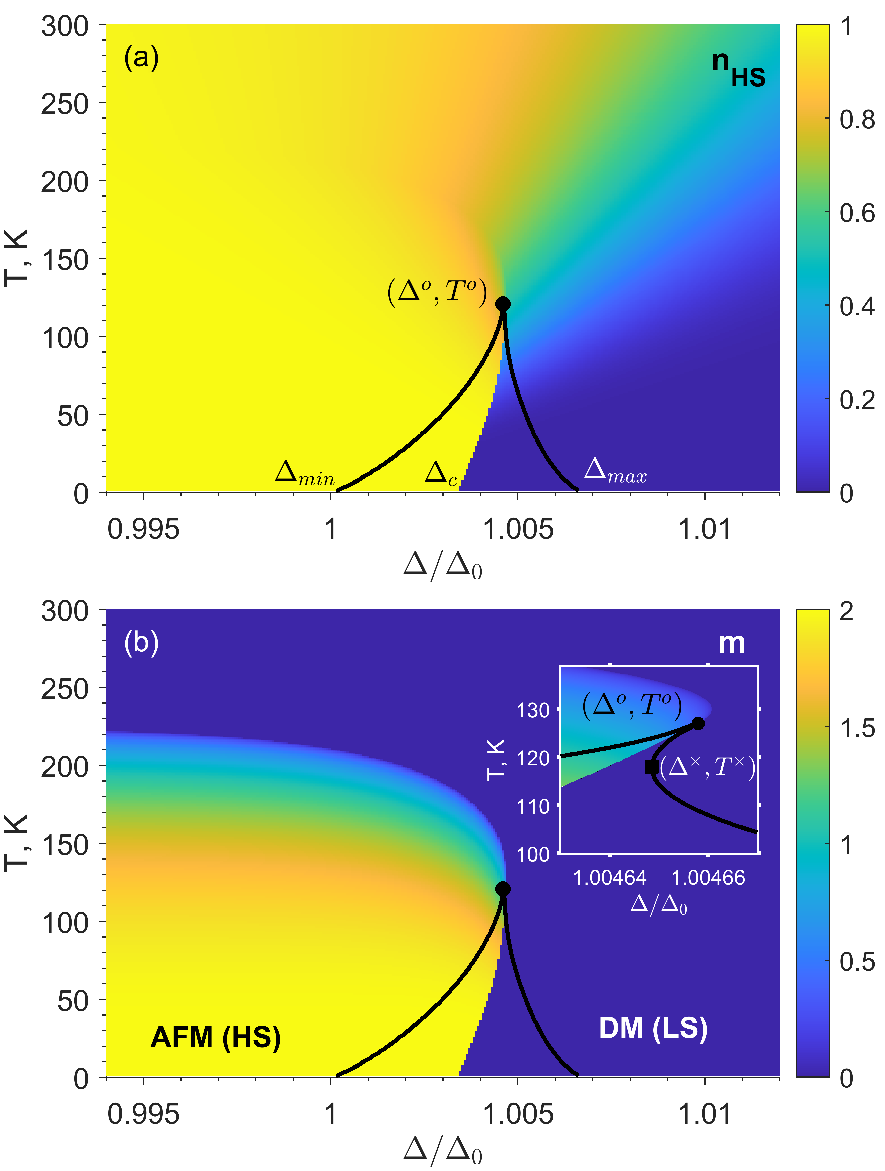}
\caption{\label{fig2} ${\Delta} - T$ phase diagrams of the HS state population ${n_{HS}}$ (a) and the sublattice magnetization $m$ (b) for ${J_\tau } = 0$ and ${J_S} = 112$~K. The inset to Fig.~\ref{fig2}b shows an enlarged view of the metastable region boundaries near the tricritical point $\left( {\Delta^\odot ,{T^\odot }} \right)$, where the inflection point $\left( {\Delta^ \times ,{T^ \times }} \right)$ of the right boundary is clearly visible. The calculations were performed with $g_\tau = 3$.}
\end{figure}

Figure~\ref{fig2} shows calculated phase diagrams of the HS state population $n_{HS}$ (Fig.~\ref{fig2}a) and the sublattice magnetization $m = \left| {{m_C}} \right|$ (Fig.~\ref{fig2}b) for ${J_S} = 112$~K and ${J_\tau } = 0$. Although in the single-ion picture for ${\Delta} > \Delta_0$ the LS state is the ground state (Fig.~\ref{fig1}a), it is seen that due to the cooperative interaction ${J_S}$, the HS state remains the ground state even for ${\Delta} > \Delta_0$, but for ${\Delta} < \Delta_C$, where $\Delta_C$ is the critical value of the crystal field at which the ground state changes (Fig.~\ref{fig2}a). The exchange interaction  ${J_S}$ stabilizes the HS state by lowering its energy, hence $\Delta_C > \Delta_0$. For ${\Delta} > \Delta_C$ the antiferromagnetic (AFM) HS ground state is replaced by the diamagnetic (DM) LS state (Fig.~\ref{fig2}b). The elastic interaction ${J_\tau }$, in contrast to the exchange interaction $J_S$, does not lead to an increase in the critical value of the crystal field ($\Delta_C = \Delta_0$, Fig.~\ref{fig1}b). The phase diagrams (Fig.~\ref{fig2}a and \ref{fig2}b) show the existence of a special tricritical point $\left( {\Delta^ \odot ,{T^ \odot }} \right)$, where the first-order phase transition line continuously changes into a second-order phase transition line. The black solid lines, as in the case considered in Fig.~\ref{fig1}b, indicate the region of metastable states.

\section{Self-oscillations of HS/LS state populations and magnetization \label{Auto_n_m}}
Consider an SC system in thermal contact with a thermostat at temperature ${T_R}$ and exposed to external radiation of intensity ${I_0}$, which leads to its photothermal heating. The time dependence $t$ of the system temperature $T$ can be described by the equation
\begin{eqnarray}
    \frac{{\partial T}}{{\partial t}} = &-& \alpha \left( {T - {T_R}} \right) \nonumber \\
    &+& I\left[ 1 + \left( {\rho - 1} \right) \left( \tau + \frac{1}{2} \right) \right]
    - \frac{{\Delta H }}{{{C_p}}}\frac{{\partial \kappa }}{{\partial t}}.
    \label{main_dT/dt}
\end{eqnarray}
Here, the first term on the right-hand side describes the coupling to the thermostat ($\alpha $ is the coupling coefficient). The second term arises from the absorption of external radiation: $I = {I_0}{\sigma _a}$, where ${\sigma _a}$ is the absorption cross-section; $\rho = {a_{HS}}/{a_{LS}}$, with $a_{HS(LS)}$ being the absorption coefficients in the HS (LS) state. The last term in \eqref{main_dT/dt}, where $\kappa = \tau$ if $J_S = 0$ and $\kappa = m$ if $J_S \ne 0$, describes the temperature change due to the change in enthalpy $\Delta H$ resulting from a first-order phase transition. Here, $\Delta H = {T_{eq}}\Delta S$, with ${T_{eq}} = \frac{{{\Delta _S}}}{{{k_B}\ln g}}$ being the spin transition temperature at which 
${n_{HS}} = {n_{LS}} = {1 \mathord{\left/ 
 {\vphantom {1 2}} \right.
 \kern-\nulldelimiterspace} 2}$ or ${\tau} = 0$, $\Delta S = R\ln g$ is the entropy change ($R$ is the molar gas constant).
\begin{eqnarray*}
    \Delta S &=& {S_{HS}} - {S_{LS}} \nonumber \\
    &=& R\ln \left( {{g_{\tau ,HS}} \cdot {g_{S,HS}}} \right) - R\ln \left( {{g_{\tau ,LS}} \cdot {g_{S,LS}}} \right) \nonumber \\
    &=& R\ln \left( {{g_\tau }} \right) + R\ln \left( {{g_S}} \right) = \Delta {S_\tau } + \Delta {S_S},
\end{eqnarray*}
Therefore, the total enthalpy change $\Delta H$ can be represented as the sum of two contributions, $\Delta {H_\tau } = T_{eq}\Delta S_\tau$ and $\Delta {H_S} = T_{eq}\Delta S_S$, associated with the changes in orbital and spin entropy.

Following the Glauber formalism~\cite{Glauber_JMathematicalPhysics.4.294.1963}, we write the equation of motion for the probability $p\left[ {\left( {{\sigma _1},{s_1}} \right), \ldots ,\left( {{\sigma _N},{s_N}} \right);t} \right]$ of finding a system of $N$  pseudospins and  $N$ spins in the state with configuration $\left\{ {\sigma ,s} \right\} = \left\{ {\left( {{\sigma _1},{s_1}} \right), \ldots ,\left( {{\sigma _N},{s_N}} \right)} \right\}$ at time  $t$ as
\begin{widetext}
\begin{eqnarray}
    &p\left[ {\left( {{\sigma _1},{s_1}} \right), \ldots ,\left( {{\sigma _N},{s_N}} \right);t} \right] = \nonumber \\
    &- \sum\limits_{i = 1}^N {\sum\limits_{{{\sigma '}_i},{{s'}_i}} {{W_i}\left[ {\left( {{\sigma _i},{s_i}} \right) \to \left( {{{\sigma '}_i},{{s'}_i}} \right)} \right]p\left[ {\left( {{\sigma _1},{s_1}} \right), \ldots ,\left( {{\sigma _i},{s_i}} \right), \ldots ,\left( {{\sigma _N},{s_N}} \right);t} \right]} } \nonumber \\
    &+ \sum\limits_{i = 1}^N {\sum\limits_{{{\sigma '}_i},{{s'}_i}} {{W_i}\left[ {\left( {{{\sigma '}_i},{{s'}_i}} \right) \to \left( {{\sigma _i},{s_i}} \right)} \right]p\left[ {\left( {{\sigma _1},{s_1}} \right), \ldots ,\left( {{{\sigma '}_i},{{s'}_i}} \right), \ldots ,\left( {{\sigma _N},{s_N}} \right);t} \right]} },
    \label{Glauber_master_equation}
\end{eqnarray}
\end{widetext}
where the coefficients ${W_i}\left[ {\left( {{\sigma _i},{s_i}} \right) \to \left( {{{\sigma '}_i},{{s'}_i}} \right)} \right]$ are the transition probabilities per unit time for the system to go from a state with configuration $\left\{ {\left( {{\sigma _1},{s_1}} \right), \ldots ,\left( {{\sigma _i},{s_i}} \right), \ldots ,\left( {{\sigma _N},{s_N}} \right)} \right\}$ to a state with configuration $\left\{ {\left( {{\sigma _1},{s_1}} \right), \ldots ,\left( {{{\sigma '}_i},{{s'}_i}} \right), \ldots ,\left( {{\sigma _N},{s_N}} \right)} \right\}$ which differs from the original one only in the state of a single site $i$: $\left( {{{\sigma '}_i},{{s'}_i}} \right) \ne \left( {{\sigma _i},{s_i}} \right)$. Here and below, to describe the state of an ion at site $i$, we use the notation $\left( {{\sigma _i},{s_i}} \right)$, where  ${s_i} =  - S, - S + 1,...,S$ if ${\sigma _i} = \frac{1}{2}$ (HS state), and ${s_i} = 0$ if ${\sigma _i} =  - \frac{1}{2}$ (LS state).

According to the detailed balance principle
\begin{eqnarray}
    &\frac{{{W_i}\left[ {\left( {{\sigma _i},{s_i}} \right) \to \left( {{{\sigma '}_i},{{s'}_i}} \right)} \right]}}{{{W_i}\left[ {\left( {{{\sigma '}_i},{{s'}_i}} \right) \to \left( {{\sigma _i},{s_i}} \right)} \right]}} = \nonumber \\
    &= \frac{{p\left[ {\left( {{\sigma _1},{s_1}} \right), \ldots ,\left( {{{\sigma '}_i},{{s'}_i}} \right), \ldots ,\left( {{\sigma _N},{s_N}} \right);t} \right]}}{{p\left[ {\left( {{\sigma _1},{s_1}} \right), \ldots ,\left( {{\sigma _i},{s_i}} \right), \ldots ,\left( {{\sigma _N},{s_N}} \right);t} \right]}}.
\end{eqnarray}
Taking into account that $p\left[ {\left\{ {\sigma ,s} \right\}} \right] \sim {e^{ - \beta {E_{\left\{ {\sigma ,s} \right\}}}}}$, where ${E_{\left\{ {\sigma ,s} \right\}}}$ is the energy of the state with configuration $\left\{ {\sigma ,s} \right\}$, we obtain
\begin{eqnarray}
    &\frac{{{W_i}\left[ {\left( {{\sigma _i},{s_i}} \right) \to \left( {{{\sigma '}_i},{{s'}_i}} \right)} \right]}}{{{W_i}\left[ {{{\left( {{\sigma '}_i,{{s'}_i}} \right)} } \to \left( {{\sigma _i},{s_i}} \right)} \right]}} = \nonumber \\
    &= \frac{{\exp \left( { - \beta E_i^\tau {{\sigma '}_i}} \right)\exp \left[ { - \beta E_i^S\left( {{{\sigma '}_i} + \frac{1}{2}} \right){{s'}_i}} \right]}}{{\exp \left( { - \beta E_i^\tau {\sigma _i}} \right)\exp \left[ { - \beta E_i^S\left( {{\sigma _i} + \frac{1}{2}} \right){s_i}} \right]}},
    \label{otnoshenie_W}
\end{eqnarray}
where $E_i^\tau  = {\Delta _\tau } - {J_\tau }\sum\limits_{\delta  = 1}^z {{\sigma _{i + \delta }}} $, and $E_i^S = {J_S}\sum\limits_{\delta  = 1}^z {\left( {{\sigma _{i + \delta }} + \frac{1}{2}} \right){s_{i + \delta }}} $.

Equation \eqref{otnoshenie_W} determines only the ratio of the transition rates; one can assume that
\begin{gather}
{W_i}\left[ {\left( {{\sigma _i},{s_i}} \right) \to \left( {{{\sigma '}_i},{{s'}_i}} \right)} \right] = \nonumber \\
= A\left( T \right)\exp \left( { - \beta E_i^\tau {{\sigma '}_i}} \right)\exp \left[ { - \beta E_i^S\left( {{{\sigma '}_i} + \frac{1}{2}} \right){{s'}_i}} \right],
\end{gather}
where the coefficient $A\left( T \right)$ depends on the temperature and on system parameters such as the spin-orbit coupling parameter $\xi  = \left\langle {{}^1{A_{1g}}\left| {{H_{SO}}} \right|{}^5{T_{2g}}} \right\rangle $ and the electron-phonon interaction parameter ${g_1}$. A sufficiently good approximation is the dependence $A\left( T \right) = \frac{1}{{{\tau _0}}}{e^{ - \beta E_a^{\left( 0 \right)}}}$, where $\tau _0^{ - 1}$ is the characteristic intrinsic frequency, and $E_a^{\left( 0 \right)}$ is the intramolecular energy barrier height due to vibronic interactions~\cite{PhysRevB.62.14796.2000}.

The average values of the random variables ${\sigma _i}\left( t \right)$ and ${s_i}\left( t \right)$ can be expressed as
\begin{eqnarray}
    \left\langle {{\eta _i}\left( t \right)} \right\rangle  = \sum\limits_{\left\{ {\sigma ,s} \right\}} {{\eta _i}p\left[ {\left\{ {\sigma ,s} \right\};t} \right]}.
    \label{eq:sigma(t)_s(t)}
\end{eqnarray}
For brevity, when writing similar equations for ${\sigma _i}\left( t \right)$ and ${s_i}\left( t \right)$ here and in Eqs.~\eqref{eq:dsigma/dt_dsdt}, \eqref{eq:dsigma/dt_dsdt (two part)}, we use the notation $\eta  = \sigma ,s$. Using the master equation \eqref{Glauber_master_equation}, from \eqref{eq:sigma(t)_s(t)} we obtain
\begin{eqnarray}
    &\frac{d}{{dt}}\left\langle {{\eta _i}\left( t \right)} \right\rangle  = \nonumber \\
    &- 2\sum\limits_{\left\{ {\sigma ,s} \right\}} {\sum\limits_{{{\sigma '}_i},{{s'}_i}} {{\eta _i}{W_i}\left[ {\left( {{\sigma _i},{s_i}} \right) \to \left( {{{\sigma '}_i},{{s'}_i}} \right)} \right]p\left[ {\left\{ {\sigma ,s} \right\};t} \right]} }  \equiv \nonumber \\
    &- 2\sum\limits_{{{\sigma '}_i},{{s'}_i}} {\left\langle {{\eta _i}{W_i}\left[ {\left( {{\sigma _i},{s_i}} \right) \to \left( {{{\sigma '}_i},{{s'}_i}} \right)} \right]} \right\rangle }.
    \label{eq:dsigma/dt_dsdt}
\end{eqnarray}
The right-hand side of Eq.~\eqref{eq:dsigma/dt_dsdt} can be represented as the sum of two contributions:
\begin{eqnarray}
    \frac{d}{{dt}}\left\langle {{\eta _i}\left( t \right)} \right\rangle = - 2\sum\limits_{{s'}_i} {\left\langle {{\eta _i}{W_i}\left[ {\left( {{\sigma _i},{s_i}} \right) \to \left( {{{\bar \sigma }_i},{{s'}_i}} \right)} \right]} \right\rangle } \nonumber \\
    - 2\sum\limits_{{{s'}_i} \ne {s_i}} {\left\langle {{\eta _i}{W_i}\left[ {\left( {\frac{1}{2},{s_i}} \right) \to \left( {\frac{1}{2},{{s'}_i}} \right)} \right]} \right\rangle },
    \label{eq:dsigma/dt_dsdt (two part)}
\end{eqnarray}
where ${\bar \sigma _i} =  - {\sigma _i}$. The first term on the right-hand side of Eq.~\eqref{eq:dsigma/dt_dsdt (two part)}  describes the change in the population $\tau  = \left\langle {{\sigma _i}\left( t \right)} \right\rangle $ and the magnetization $m = \left\langle {{s_i}\left( t \right)} \right\rangle $ due to pseudospin flips ($LS \mathbin{\lower.3ex\hbox{$\buildrel\textstyle\rightarrow\over
{\smash{\leftarrow}\vphantom{_{\vbox to.5ex{\vss}}}}$}} HS$ transitions). The second term is an orientational term that describes the change in magnetization caused by a change in the spatial orientation (ordering/disordering) of spins in the HS state. Interestingly, the orientational term leads to a change in the HS state population when the magnetization changes. This will be shown more explicitly below.

In the mean-field approximation, Eqs.~\eqref{eq:dsigma/dt_dsdt (two part)} can be reduced to the form
\begin{eqnarray}
\begin{split}
    &\frac{d}{{dt}}{n_{HS}} = \chi_1 \left( {{n_{HS}},m} \right), \\
    &\frac{d}{{dt}}m = \chi_2 \left( {{n_{HS}},m} \right),
\end{split}
\end{eqnarray}
where $\chi_{1\left(2\right)} \left( {{n_{HS}},m} \right)$ are certain functions, or to the form
\begin{eqnarray}
\begin{split}
    &\frac{d}{{dt}}{n_{HS}} =  - \frac{d}{{d{n_{HS}}}}U\left( {{n_{HS}},m} \right), \\
    &\frac{d}{{dt}}m =  - \frac{d}{{dm}}U\left( {{n_{HS}},m} \right),
\end{split}
\label{dn/dt_dm/dt (dynamical potential)}
\end{eqnarray}
where
$$
U\left( {{n_{HS}},m} \right) =  - \int\limits_0^{{n_{HS}}} {{\chi _1}\left( {x,m} \right)dx}  =  - \int\limits_0^m {{\chi _2}\left( {{n_{HS}},x} \right)dx} 
$$
is a dynamical potential. If we take the thermodynamic Helmholtz potential (free energy $F$) as $U$, then instead of \eqref{dn/dt_dm/dt (dynamical potential)} we obtain the Landau–Khalatnikov relaxation equations:
\begin{subequations} \label{main_(dtau/dt)_(dm/dt)}
\begin{flalign}
    \label{main_dm/dt}
    &\frac{\partial m}{\partial t} = \Gamma {J_S} n_{HS} \biggl( \frac{\theta}{e^{\beta D_\tau } + \theta}S{B_S}\left( \beta S{D_S} \right) - n_{HS}m \biggl), \\
    \label{main_dtau/dt}
    &\frac{{\partial \tau }}{{\partial t}} = \Gamma {J_\tau }\left( {\frac{1}{2}\frac{{\theta - {e^{\beta {D_\tau }}}}}{{{e^{\beta {D_\tau }}} + \theta}} } - \tau \right ) + \frac{m}{{{n_{HS}}}}\frac{{\partial m}}{{\partial t}},
\end{flalign}
\end{subequations}
where ${\Gamma}$ is the kinetic coefficient, $D_S = {{J_S}{n_{HS}}m}$, ${D_\tau } = {\Delta _\tau } - {J_\tau }\tau $,
$$
\theta  = \frac{{\sinh \left[ {\left( {2S + 1} \right)\beta {{{D_S}} \mathord{\left/
 {\vphantom {{{D_S}} 2}} \right.
 \kern-\nulldelimiterspace} 2}} \right]}}{{\sinh \left[ {\beta {{{D_S}} \mathord{\left/
 {\vphantom {{{D_S}} 2}} \right.
 \kern-\nulldelimiterspace} 2}} \right]}},
$$
$$
{B_S}\left( x \right) = \frac{{\left( {2S + 1} \right)}}{{2S}}\coth \left[ {\frac{{\left( {2S + 1} \right)}}{{2S}}x} \right] - \frac{1}{{2S}}\coth \left( {\frac{1}{{2S}}x} \right)
$$
is the Brillouin function.

From Eqs.~\eqref{main_dT/dt} and \eqref{main_(dtau/dt)_(dm/dt)} it is seen that the external radiation $I$ and the enthalpy change $\Delta H$ provide a feedback coupling between the variations in temperature $T$, state population $n_{HS}$ and magnetization $m$.

In the stationary case and in the absence of external radiation, the solutions of the self-consistent system of Eqs.~\eqref{main_dT/dt} and \eqref{main_(dtau/dt)_(dm/dt)} are thermodynamically equilibrium. Therefore, $m\left( {\Delta,T} \right)$ and ${n_{HS}}\left( {\Delta,T} \right)$, satisfying Eqs.~\eqref{main_dT/dt} and \eqref{main_(dtau/dt)_(dm/dt)} with $\frac{{\partial m}}{{\partial t}} = \frac{{\partial \tau }}{{\partial t}} = \frac{{\partial T }}{{\partial t}} = 0$ and $I=0$ coincide completely with those found by solving the self-consistent eigenvalue problem~\eqref{eig}. 

From Eq.~\eqref{main_dm/dt} it is seen that a change in the magnetization $m$ can occur both due to a change in the population of the magnetic HS state $n_{HS}$ and due to the competition between interatomic exchange ordering and thermal disordering of spins, an orientational mechanism described by the Brillouin function. The fact that a change in population entails a change in magnetization is quite obvious; much less obvious is that a change in magnetization can lead to a change in population. The second term on the right-hand side of Eq.~\eqref{main_dtau/dt}, as can be seen, leads to the fact that a change in population occurs not only due to LS$ \mathbin{\lower.3ex\hbox{$\buildrel\textstyle\rightarrow\over
{\smash{\leftarrow}\vphantom{_{\vbox to.5ex{\vss}}}}$}} $HS transitions, but is also possible due to a change in magnetization.

In the absence of interatomic exchange interaction, $m = 0$, so instead of the two equations~\eqref{main_(dtau/dt)_(dm/dt)} we obtain a single equation~\eqref{main_dtau/dt_Js0} for $\tau$
\begin{eqnarray}
    \frac{{\partial \tau }}{{\partial t}} = - \Gamma {J_\tau }\left[ {\tau  + \frac{1}{2}\tanh \left( \beta \frac{1}{2} D_\tau \right)} \right].
    \label{main_dtau/dt_Js0}
\end{eqnarray}
The system of autonomous dynamical equations~\eqref{main_dT/dt} and~\eqref{main_dtau/dt_Js0} was first presented in paper~\cite{PhysRevB.89.224303.2014}, where its solutions in the form of limit cycles were found, describing stable self-oscillations of the population $n_{HS}$ and temperature $T$. In their calculations, the authors of Ref.~\cite{PhysRevB.89.224303.2014} used parameters characteristic of the [Fe(NCSe)(py)$_{22}$(m-bpypz)], in which self-oscillations of the HS/LS state populations, manifested as a change in the color of the sample, were first experimentally observed~\cite{AdvancedTheorySimulations.1.1800080.2018}. In this case, the contribution of $\frac{{{\Delta _H}}}{{{C_p}}}$ in Eq.~\eqref{main_dT/dt} was considered small compared to the photothermal heating $I$ and was neglected.

The solutions $T_0$ and $\tau_0$, satisfying Eqs.~\eqref{main_dT/dt} and~\eqref{main_dtau/dt_Js0} with $\frac{{\partial \tau }}{{\partial t}} = \frac{{\partial T}}{{\partial t}} = 0$, are the stationary states of the SC system:
\begin{subequations} \label{stationary_tau0_T0_JS0}
\begin{eqnarray}
    \label{stationary_tau0_JS0}
    &&{\tau _0} = - \frac{1}{2}\tanh \left( {\beta _0}\frac{1}{2}D_{\tau0} \right), \\
    \label{stationary_T0_JS0}
    &&{T_0} = {T_R} + \frac{I}{{2\alpha }}\left[ {\rho  + 1 + 2{\tau _0}\left( {\rho  - 1} \right)} \right],
\end{eqnarray}
\end{subequations}
where ${\beta _0} = {\left( {{k_B}{T_0}} \right)^{ - 1}}$, ${D_{\tau 0}} = {D_\tau }\left( {{\tau _0},{T_0}} \right)$. Obviously, ${T_0} > {T_R}$.

To describe the nonequilibrium nonlinear properties of the system of Eqs.~\eqref{main_dT/dt} and~\eqref{main_dtau/dt_Js0} with $\frac{{{\Delta _H}}}{{{C_p}}} = 0$ let us first consider the propagation of small perturbations $\delta \tau \left( t \right)$ and $\delta T\left( t \right)$. In the steady state, $\tau = {\tau _0}$ and $T = {T_0}$. After a perturbation arises, a wave propagates in the medium with oscillations of the population $\tau = {\tau _0} + \delta \tau \left( t \right)$ and temperature  $T = {T_0} + \delta T\left( t \right)$. We linearize the system of dynamical equations~\eqref{main_dT/dt} and~\eqref{main_dtau/dt_Js0} around the stationary point $\left( {{T_0},{\tau _0}} \right)$. To this end, we expand the right-hand sides of Eqs.~\eqref{main_dT/dt} and~\eqref{main_dtau/dt_Js0} in a power series in the small quantities $\delta \tau \left( t \right)$ and $\delta T\left( t \right)$ and keep only the linear terms. The linearized system of Eqs.~\eqref{main_dT/dt} and~\eqref{main_dtau/dt_Js0} can be written as
\begin{eqnarray}
    \frac{{d \bm \delta }}{{dt}} = \Lambda \bm \delta,
    \label{Lin_eq}
\end{eqnarray}
where $\bm\delta  = \left( {\begin{array}{*{20}{c}}
{\delta \tau \left( t \right)}\\
{\delta T\left( t \right)}
\end{array}} \right)$, and 
\begin{eqnarray}
\Lambda  = \left( {\begin{array}{*{20}{c}}
{{\Lambda _{11}}}&{{\Lambda _{12}}}\\
{{\Lambda _{21}}}&{{\Lambda _{22}}}
\end{array}} \right) = \left( {\begin{array}{*{20}{c}}
{{{\left. {\frac{{\partial f}}{{\partial \tau }}} \right|}_0}}&{{{\left. {\frac{{\partial f}}{{\partial T}}} \right|}_0}}\\
{{{\left. {\frac{{\partial g}}{{\partial \tau }}} \right|}_0}}&{{{\left. {\frac{{\partial g}}{{\partial T}}} \right|}_0}}
\end{array}} \right)
\label{Lambda}
\end{eqnarray}
is the Jacobian matrix, with
$$
f\left( {\tau ,T} \right) = - \Gamma {J_\tau }\left[ {\tau  + \frac{1}{2}\tanh \left( \beta \frac{1}{2} D_\tau \right)} \right],
$$
$$
g\left( {\tau ,T} \right) =  - \alpha \left( {T - {T_R}} \right) + I\left[ 1 + \left( {\rho - 1} \right) \left( \tau + \frac{1}{2} \right) \right].
$$
The subscript "0"\ in Eq.~\eqref{Lambda} denotes the derivatives evaluated at the stationary point, which it is convenient to choose as ${T_0} = {T_{eq}}$ and $\tau _0 = 0$. According to Eq.~\eqref{stationary_T0_JS0} this choice corresponds to the pumping  $I = {I_{eq}} = \frac{{2\alpha }}{{\rho  + 1}}\left( {{T_{eq}} - {T_R}} \right)$. Then
\begin{eqnarray}
\begin{split}
    &{\left. {\frac{{\partial f}}{{\partial \tau }}} \right|_0} = \Gamma {J_\tau }\left( {{\beta _{eq}}\frac{1}{4}{J_\tau } - 1} \right),
    {\left. {\frac{{\partial f}}{{\partial T}}} \right|_0} = \frac{{\Gamma {J_\tau }\ln g}}{{4{T_{eq}}}}, \\
    &{\left. {\frac{{\partial g}}{{\partial \tau }}} \right|_0} =  - 2\alpha \gamma \left( {{T_{eq}} - {T_R}} \right),
    {\left. {\frac{{\partial g}}{{\partial T}}} \right|_0} =  - \alpha,
\end{split}
\label{Lambda_ij_1}
\end{eqnarray}
where $\gamma = \frac{1-\rho}{1+\rho}$.

\begin{figure}
\centering
\includegraphics[width=9.0cm]{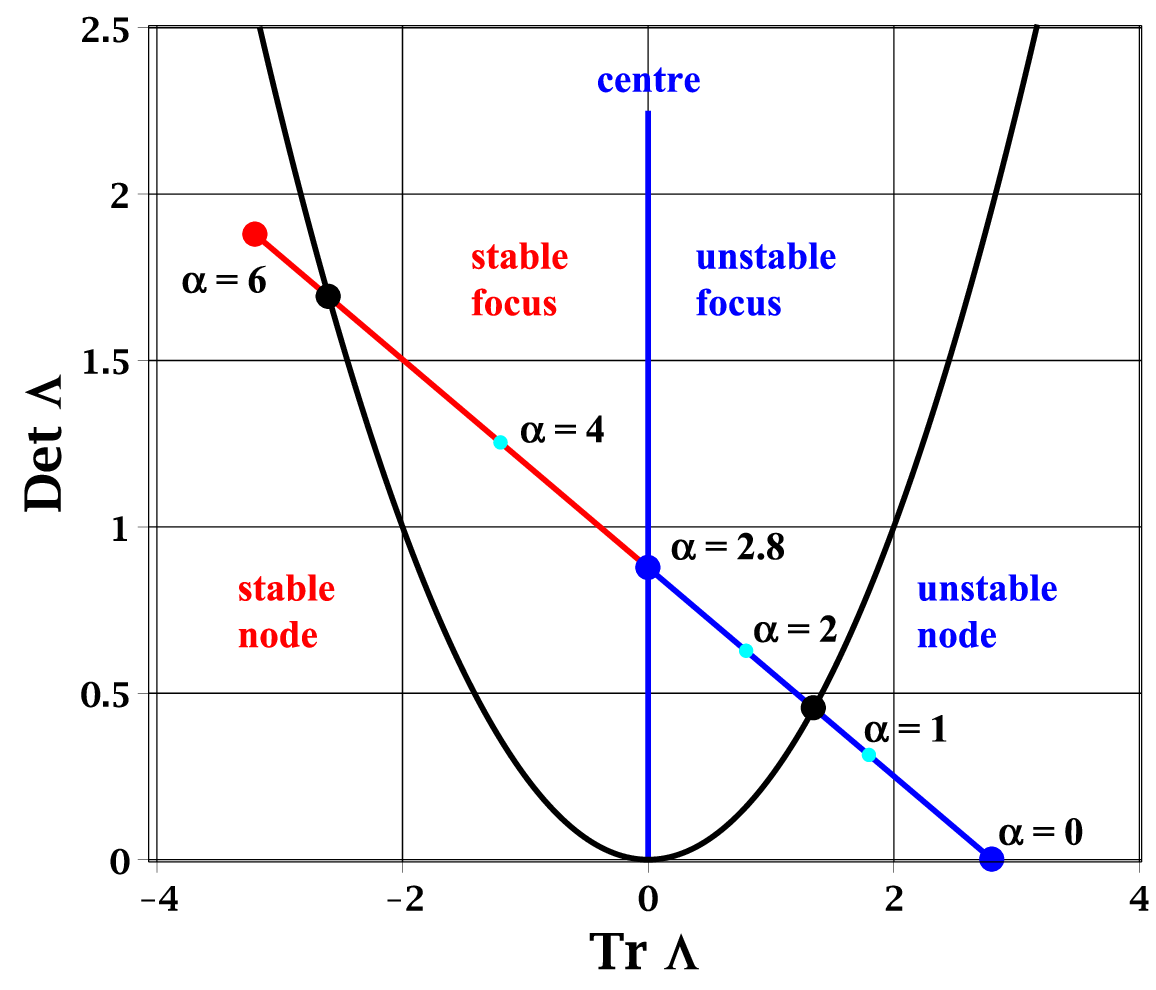}
\caption{\label{fig3} Bifurcation diagram.}
\end{figure}

\begin{figure*}
\centering
\includegraphics[width=0.30\textwidth]{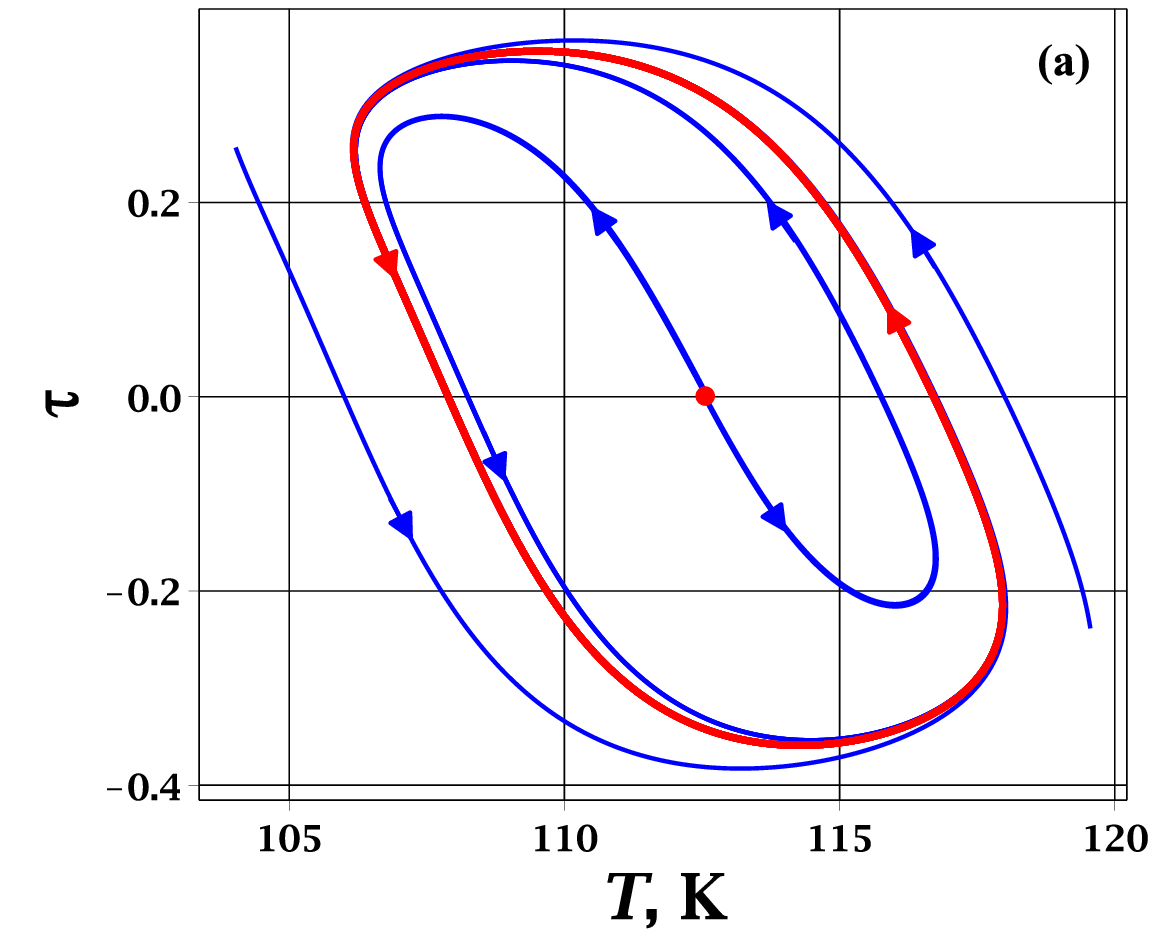}
\hfill 
\includegraphics[width=0.30\textwidth]{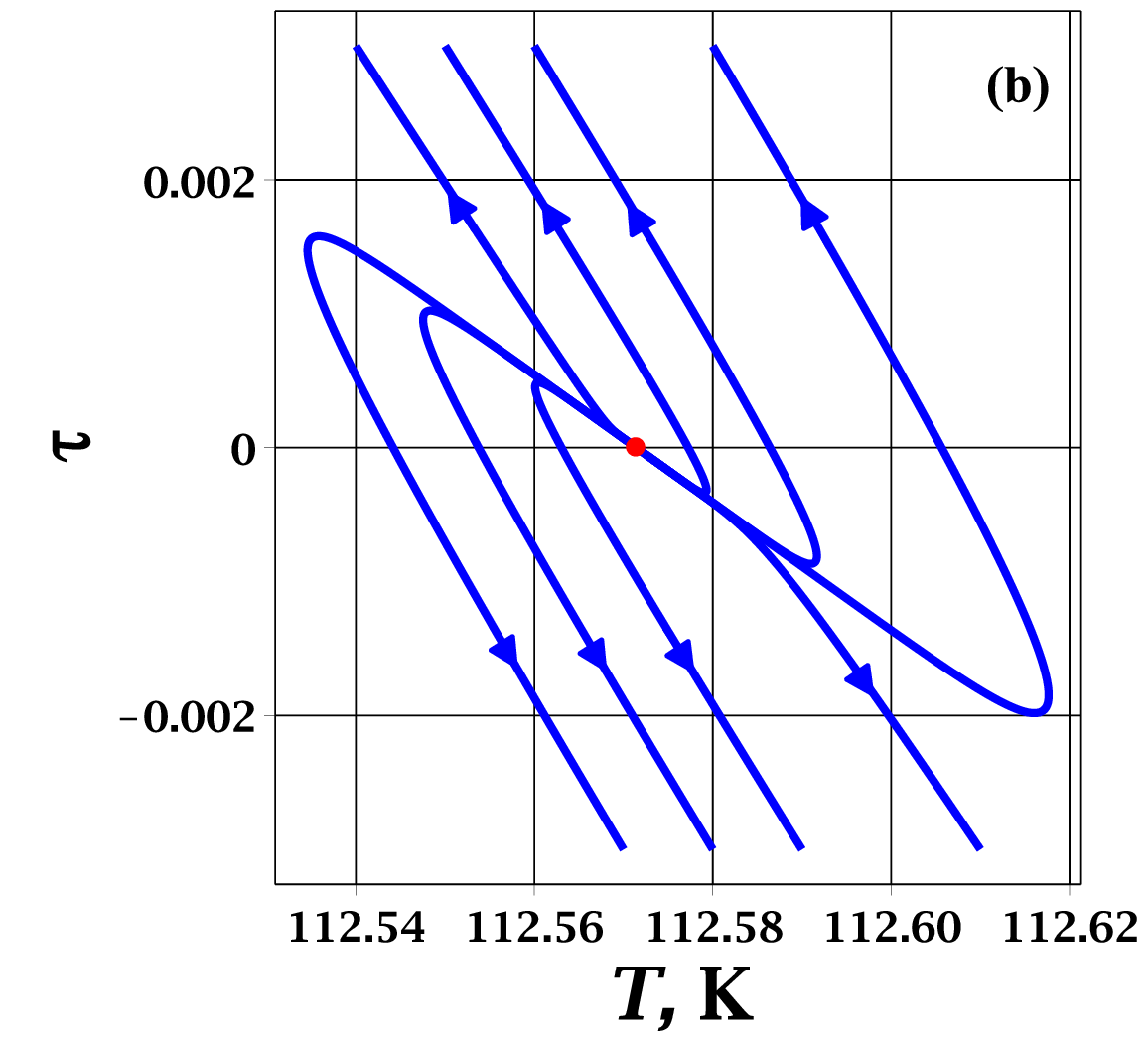}
\hfill 
\includegraphics[width=0.30\textwidth]{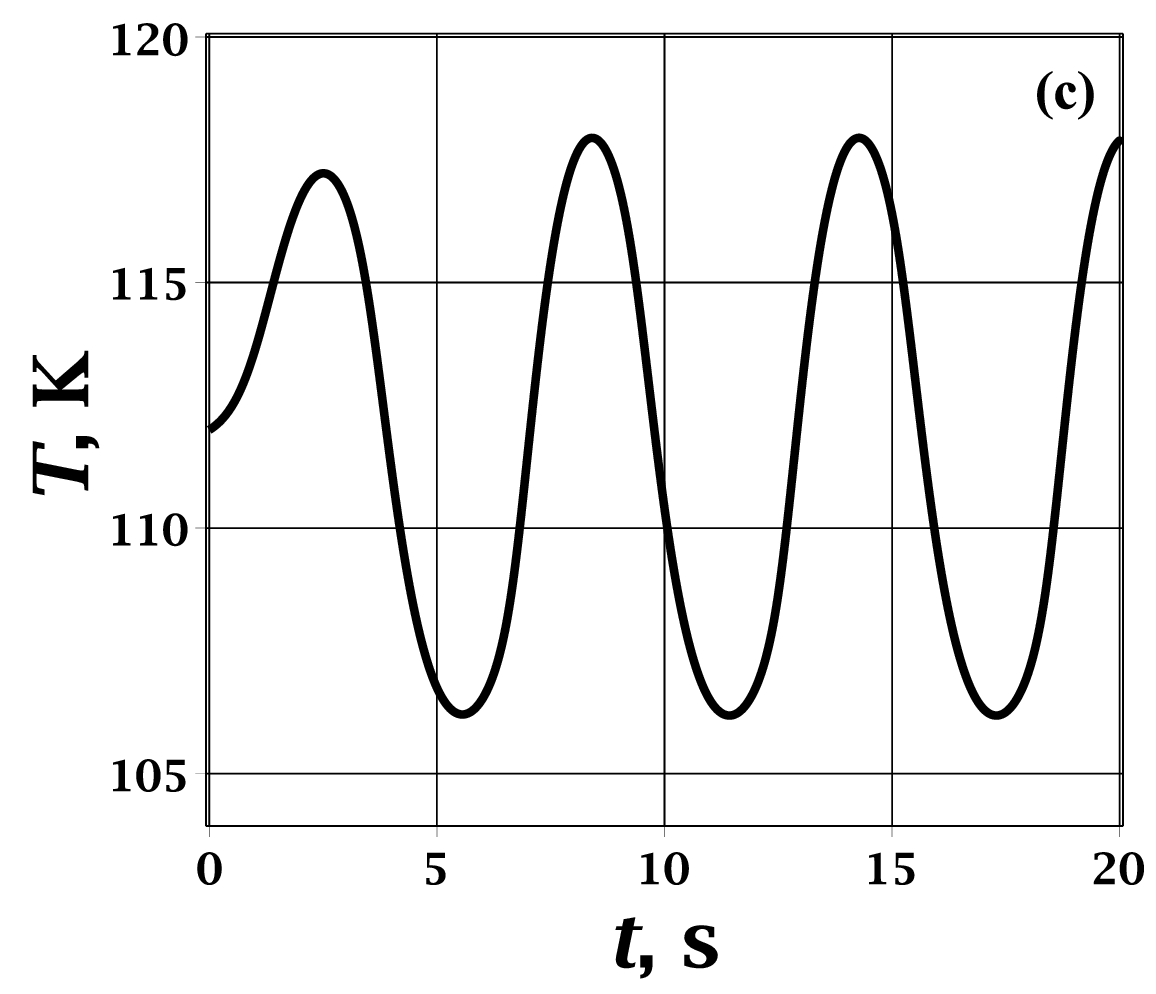}
\\ 
\includegraphics[width=0.30\textwidth]{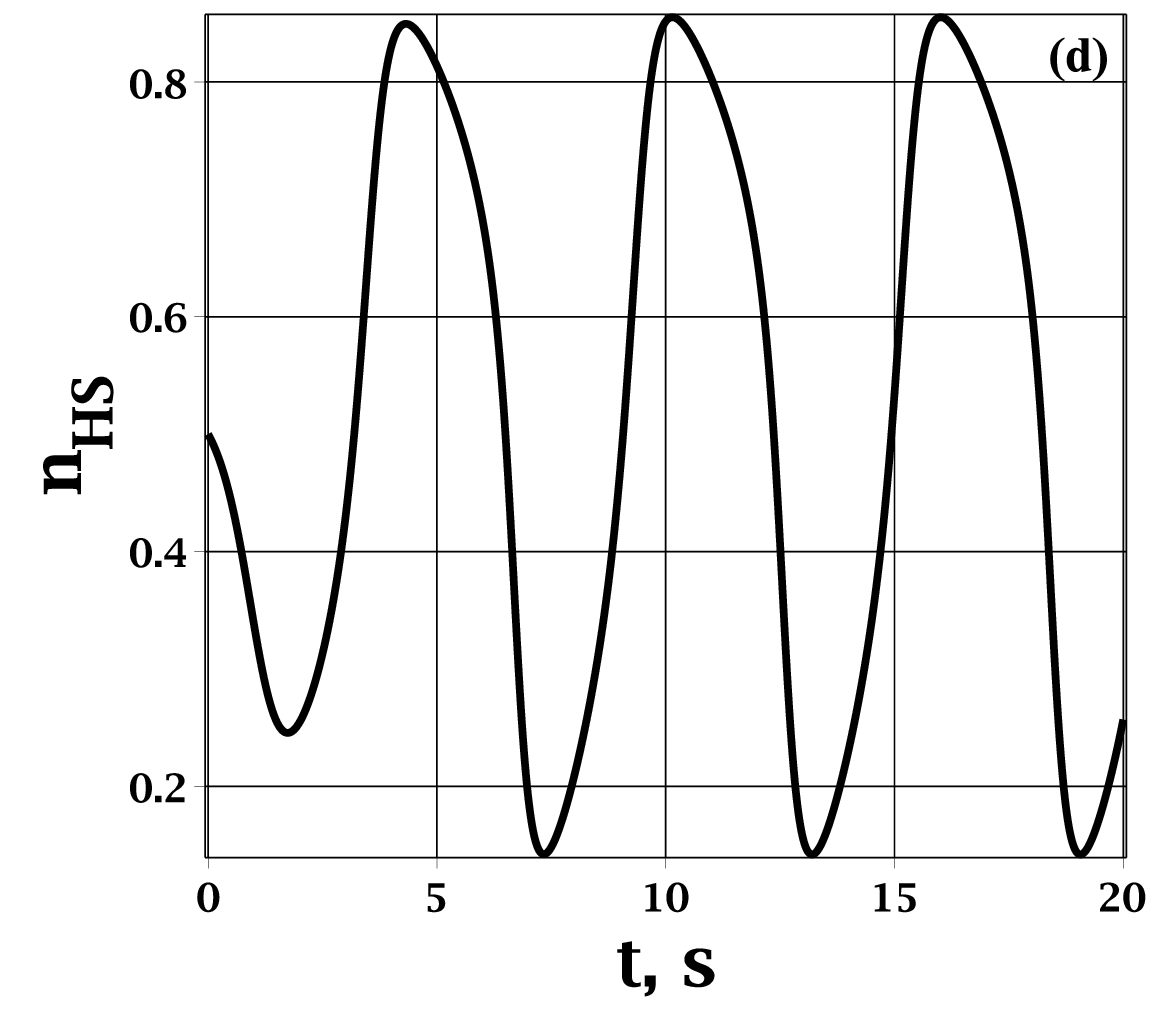}
\hfill 
\includegraphics[width=0.30\textwidth]{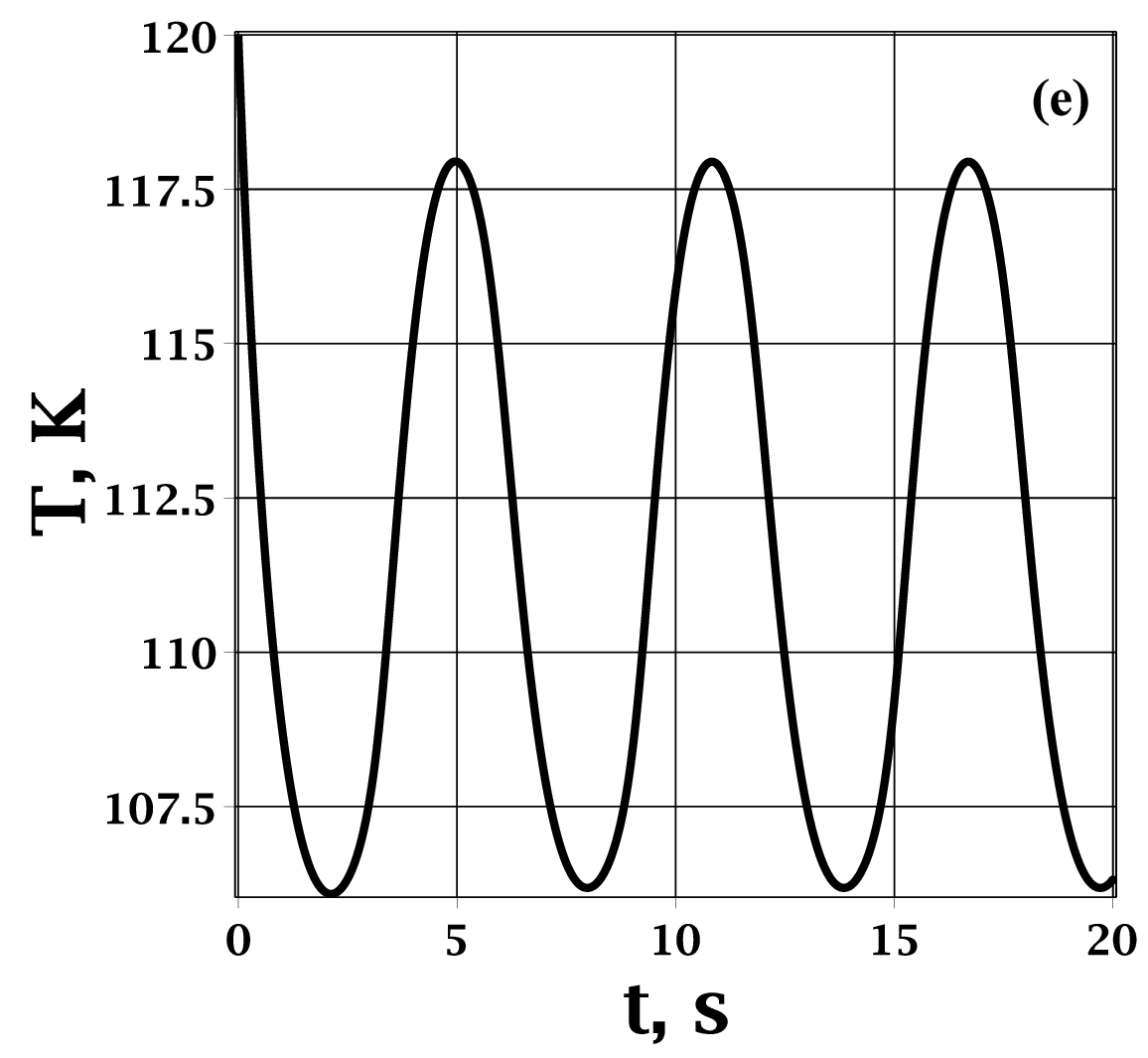}
\hfill 
\includegraphics[width=0.30\textwidth]{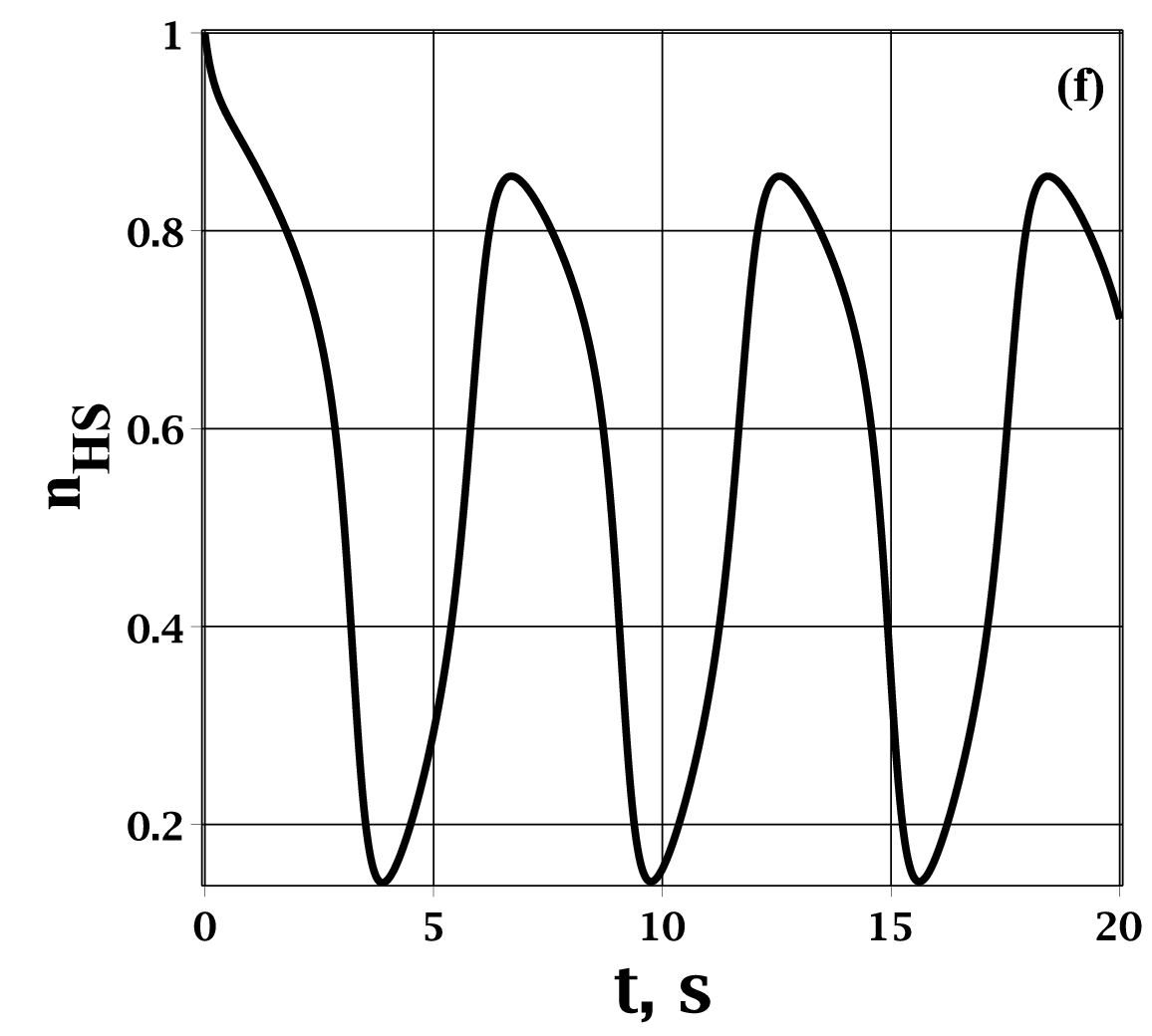}
\caption{\label{fig4} $\alpha = 1$~s$^{-1}$, unstable node. (a) Global phase portrait. The red line indicates the limit cycle; (b) Local phase portrait. The behavior of $T\left( t \right)$ and ${n_{HS}}\left( t \right)$ for initial conditions chosen inside and outside the limit cycle is shown in (c,d) and (e,f), respectively.}
\end{figure*}

\begin{figure*}
\centering
\includegraphics[width=0.30\textwidth]{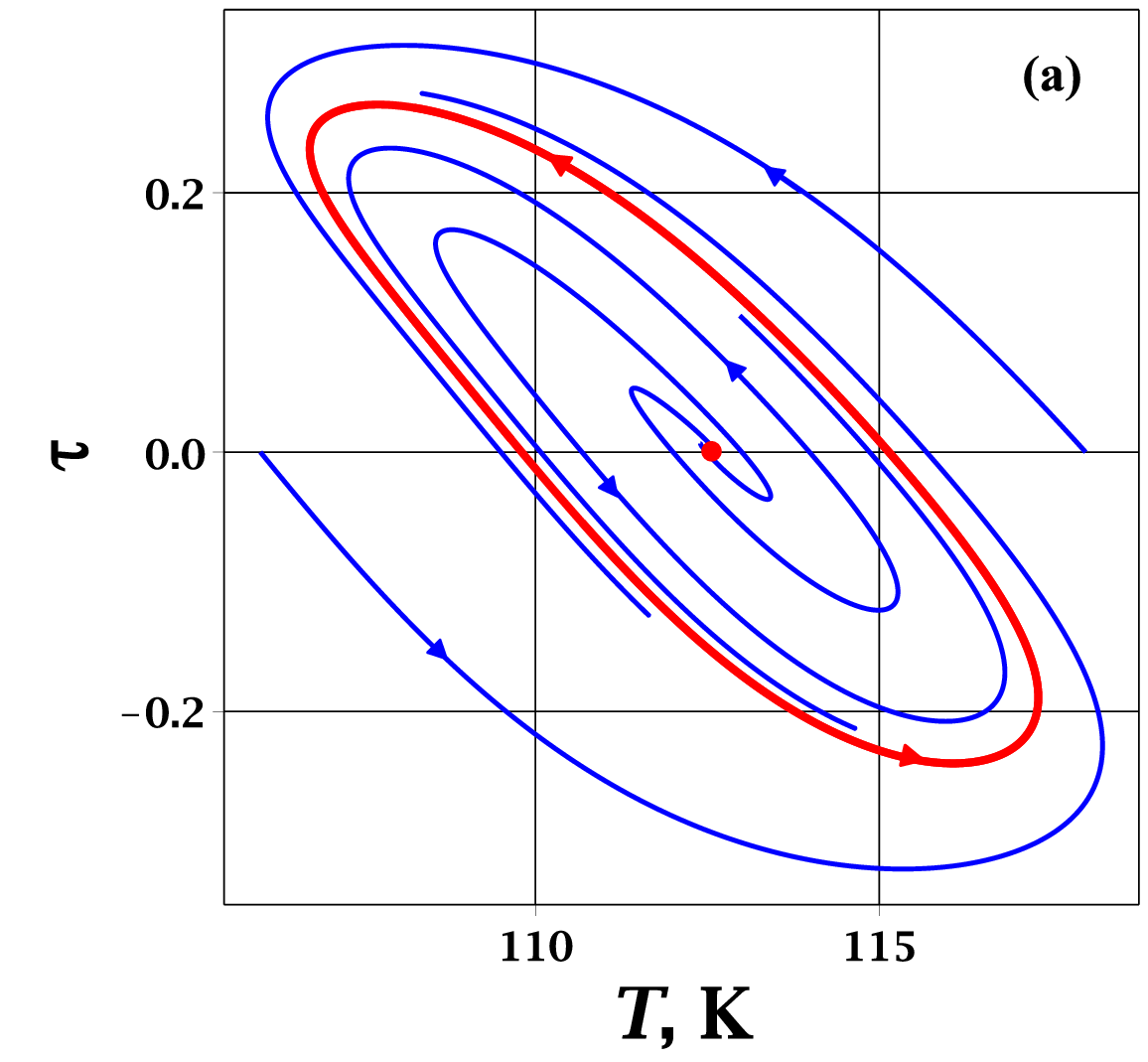}
\hfill 
\includegraphics[width=0.30\textwidth]{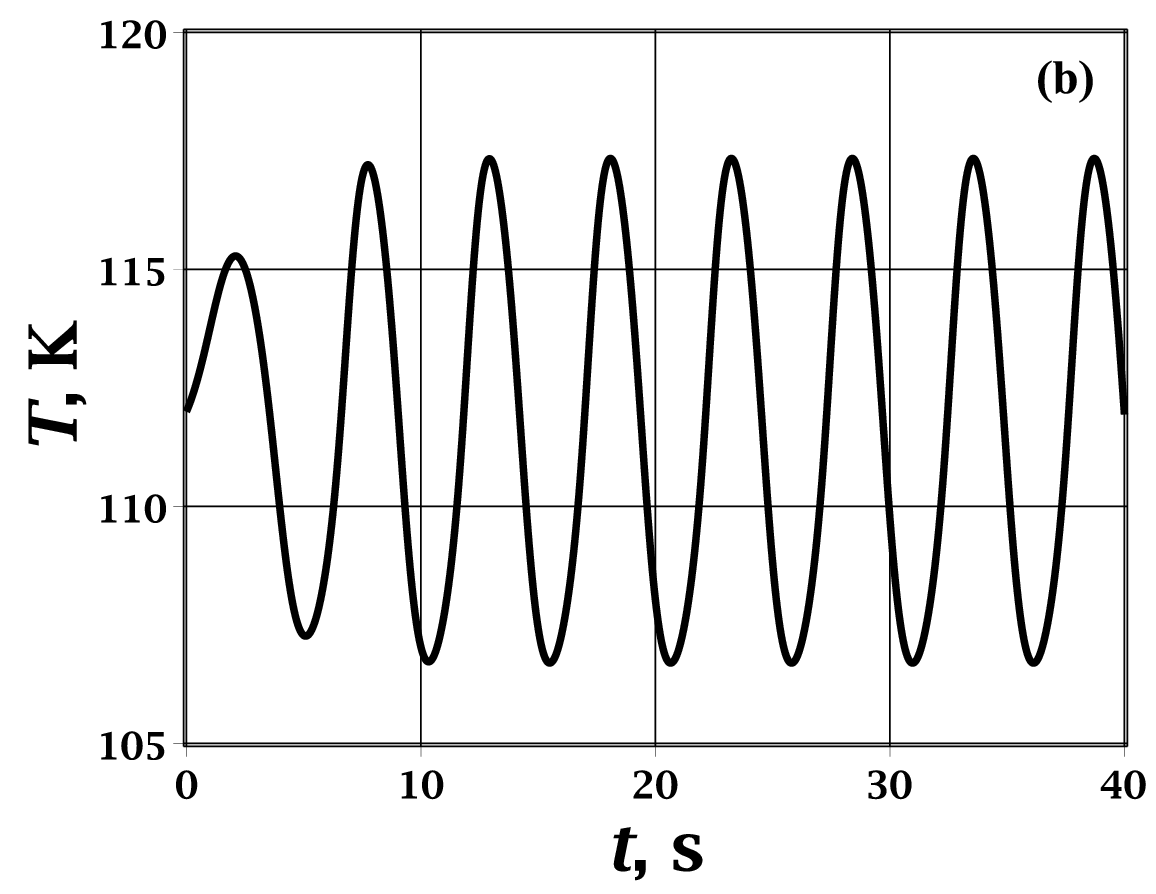}
\hfill 
\includegraphics[width=0.30\textwidth]{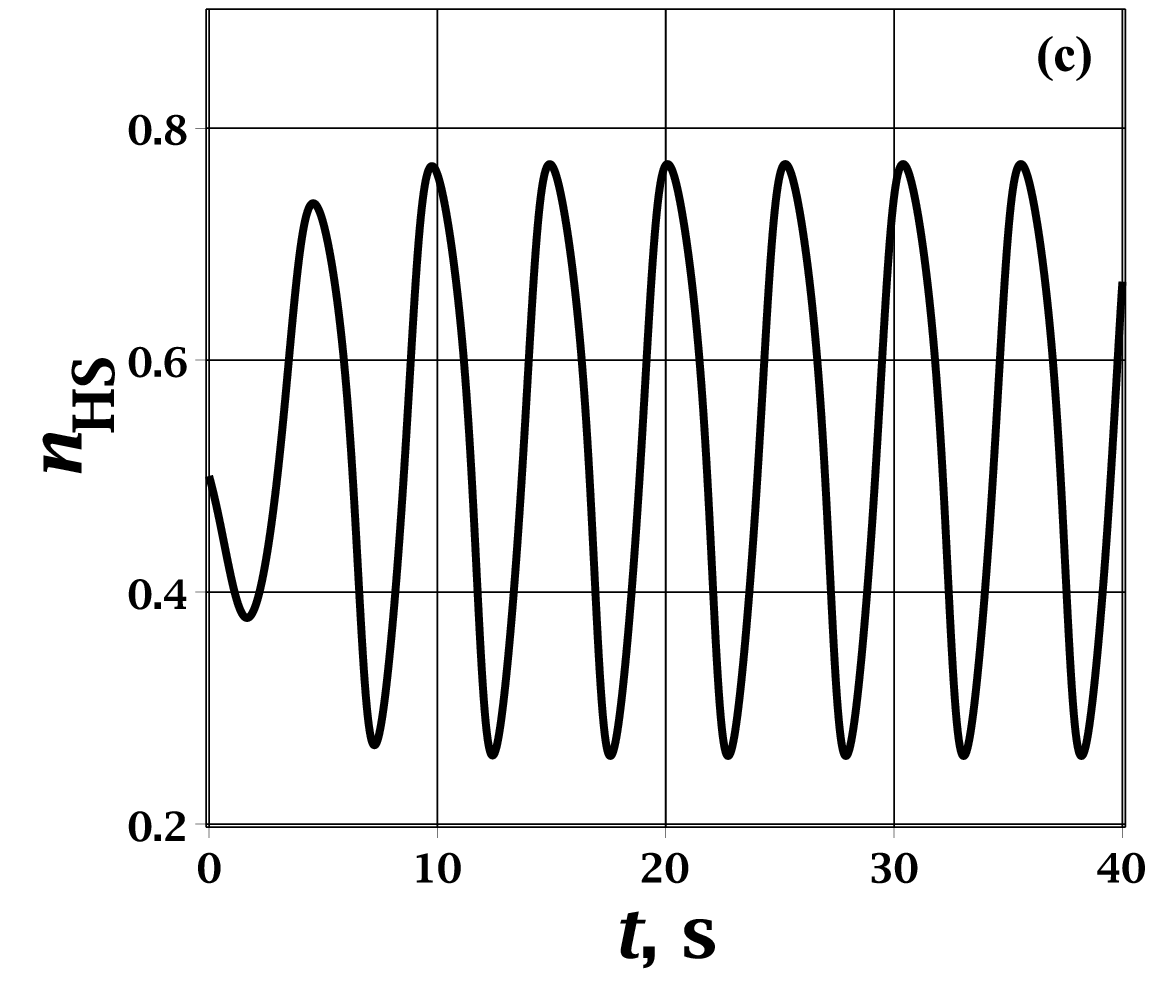}
\\ 
\hfill 
\hfill 
\hfill 
\hfill 
\hfill 
\hfill 
\hfill 
\hfill 
\includegraphics[width=0.30\textwidth]{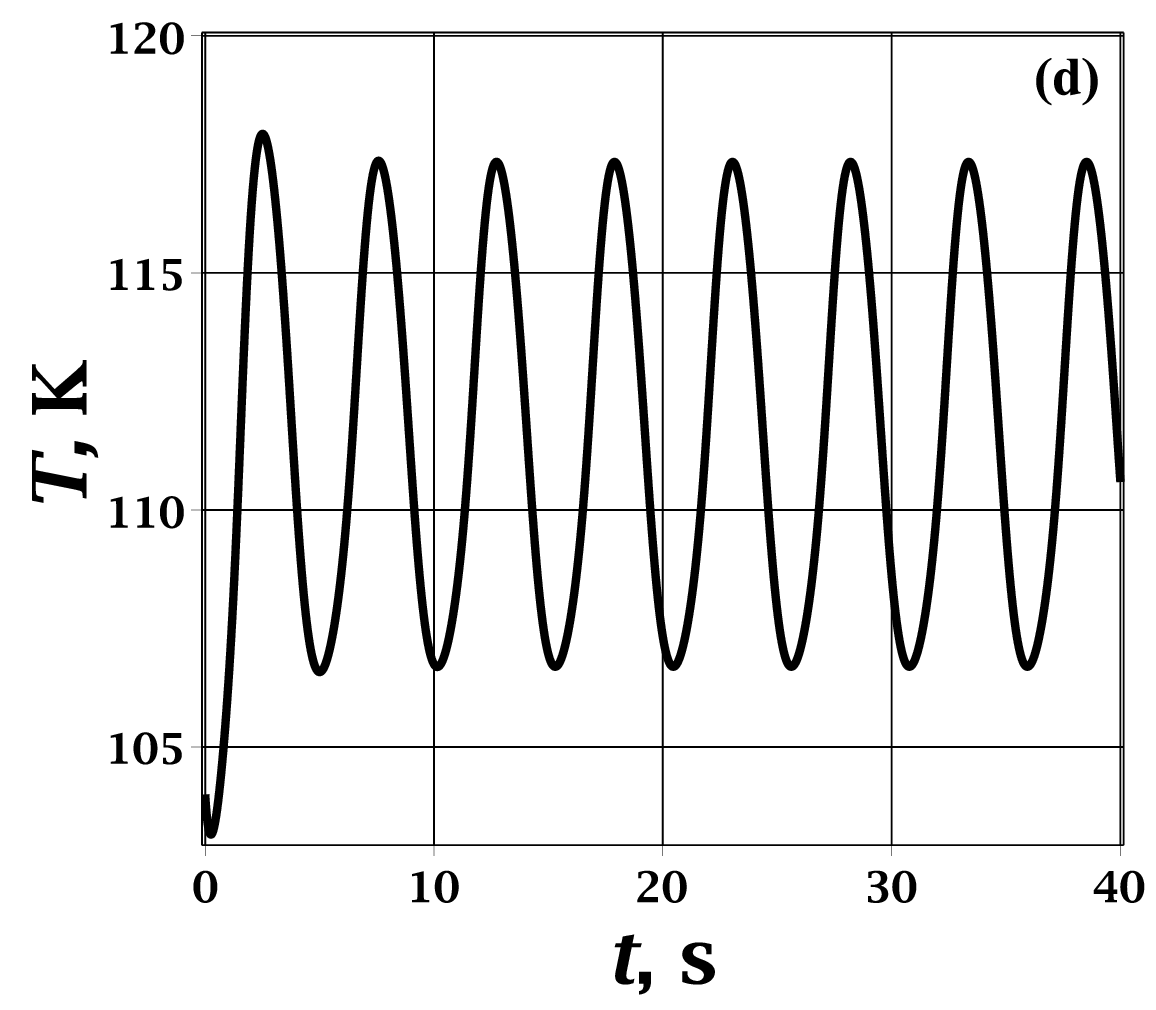}
\hfill 
\includegraphics[width=0.30\textwidth]{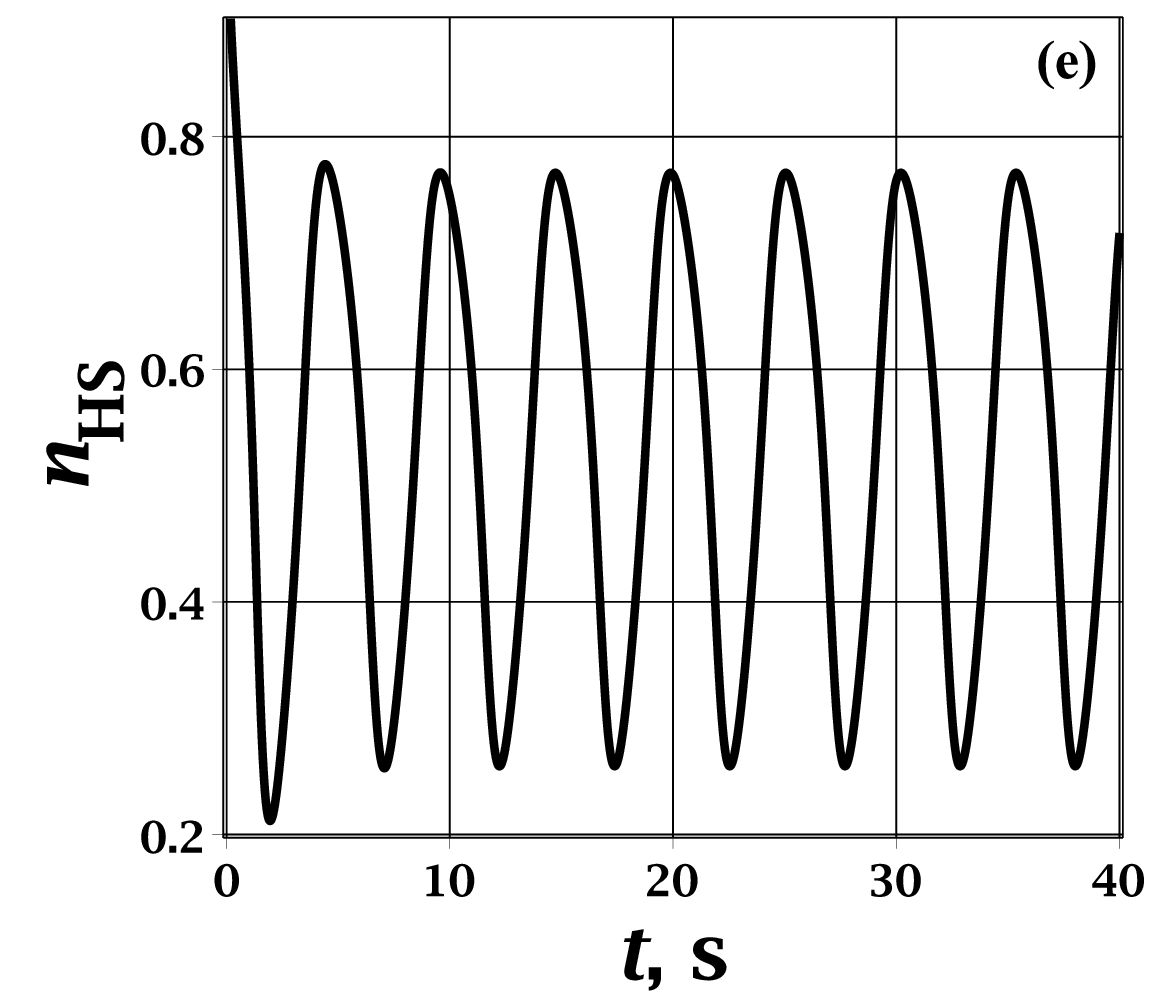}
\caption{\label{fig5} $\alpha = 2$~s$^{-1}$, unstable focus. (a) Phase portrait. The red line indicates the limit cycle; Figs.~(b) and~(c) show how, at time $t = 0$, the system leaves the initial position located inside the limit cycle and, after some time, arrives at the limit cycle. Figs.~(d) and~(e) show similar behavior, but with an initial condition outside the limit cycle.}
\end{figure*}

\begin{figure*}
\centering
\includegraphics[width=0.45\textwidth]{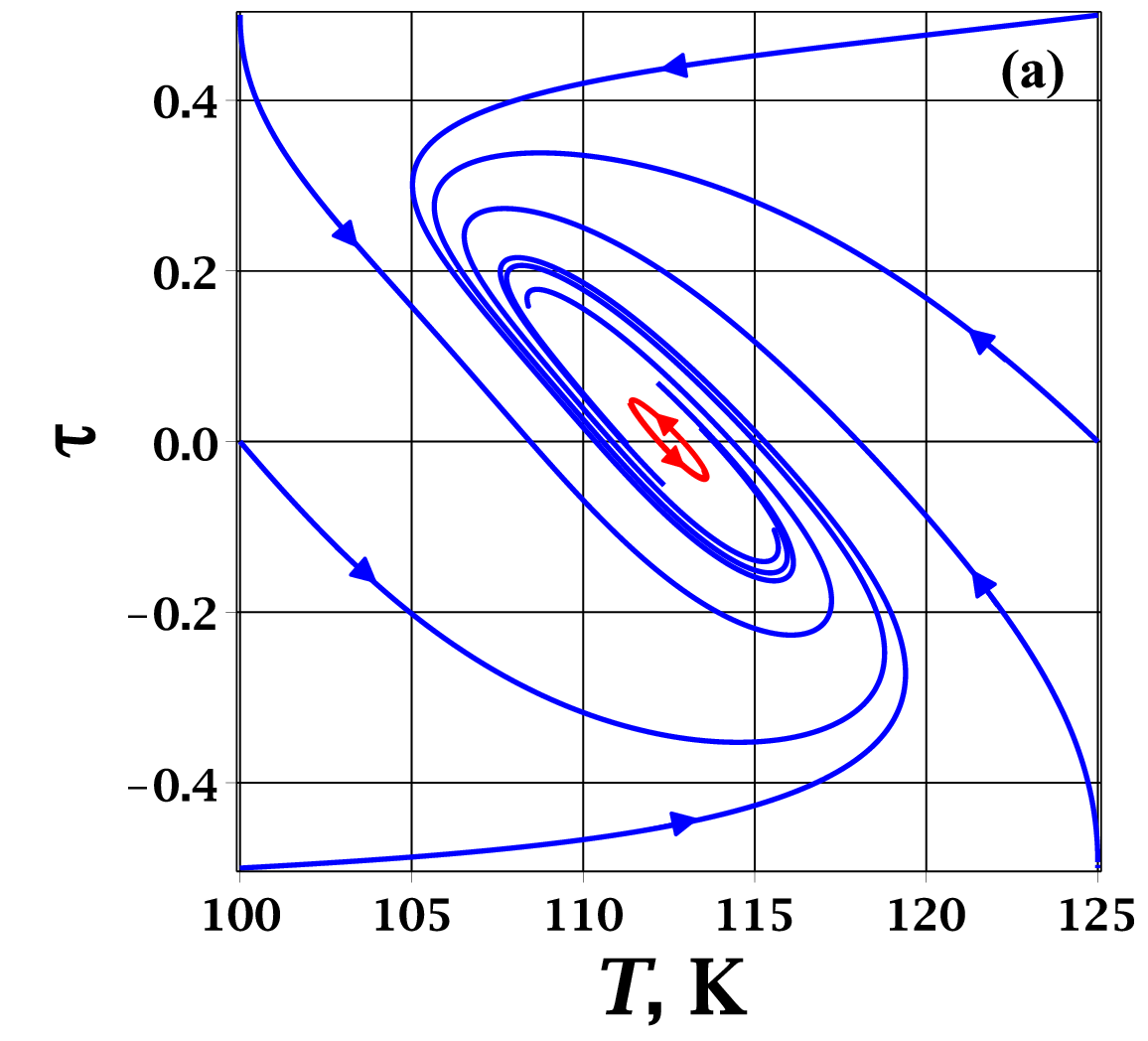}
\hfill 
\includegraphics[width=0.45\textwidth]{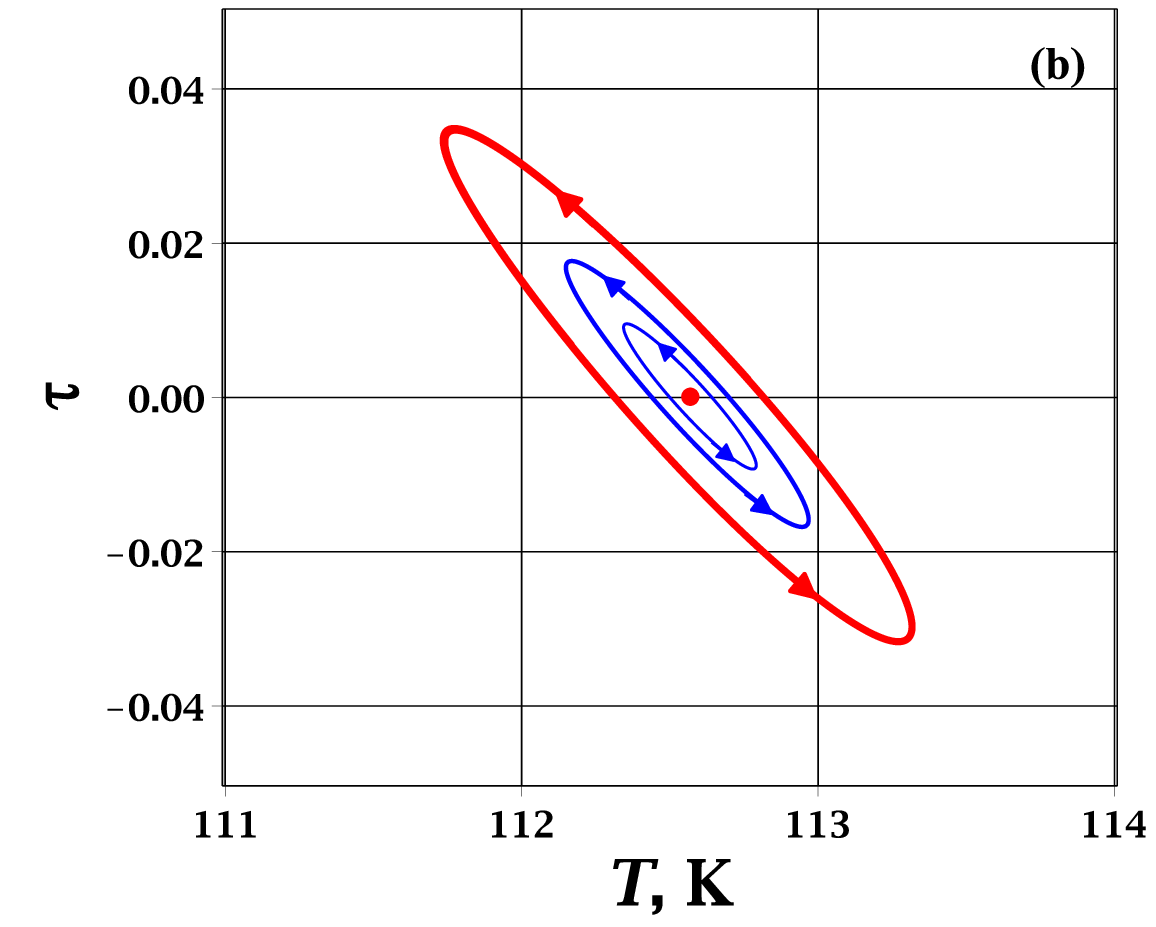}
\\ 
\includegraphics[width=0.45\textwidth]{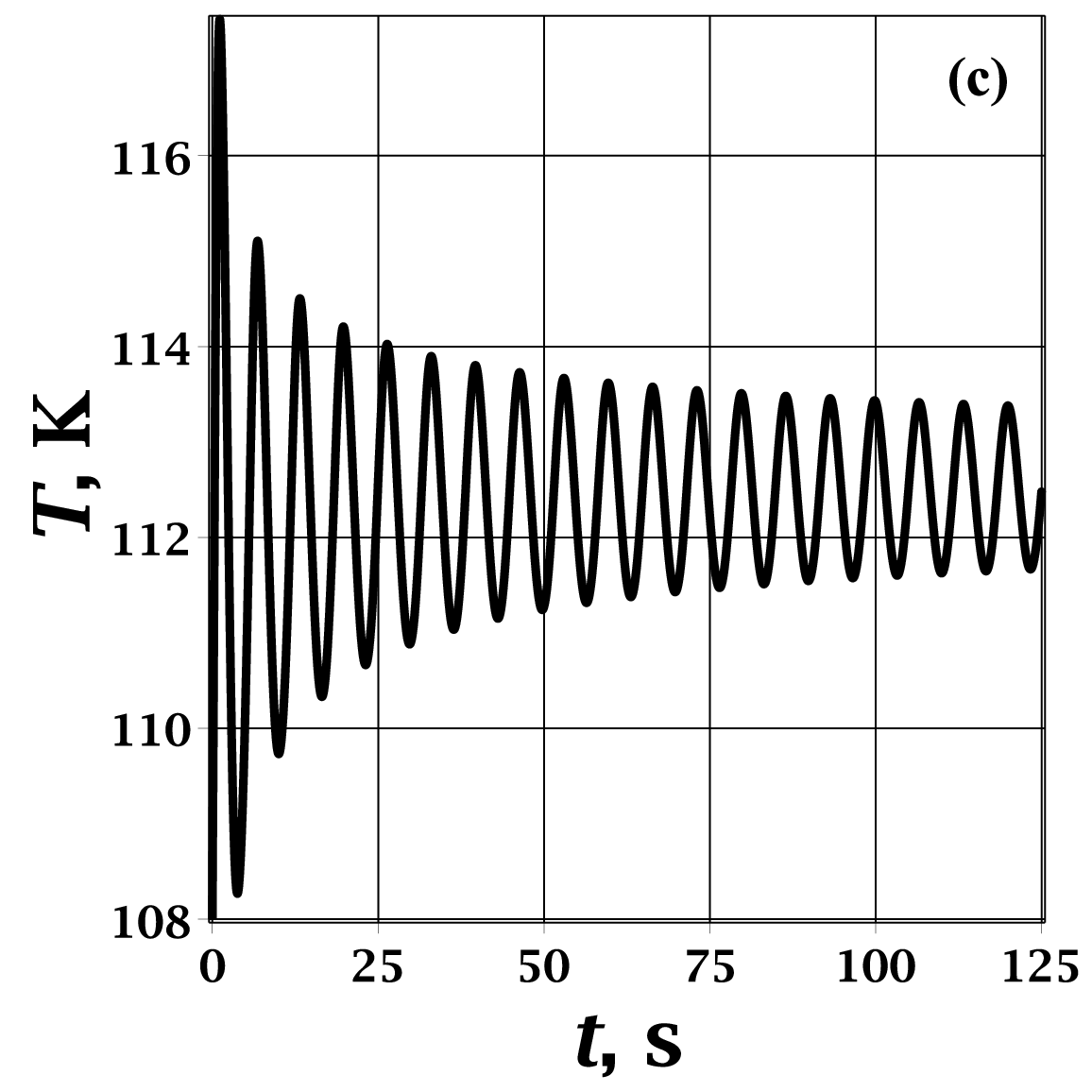}
\hfill 
\includegraphics[width=0.45\textwidth]{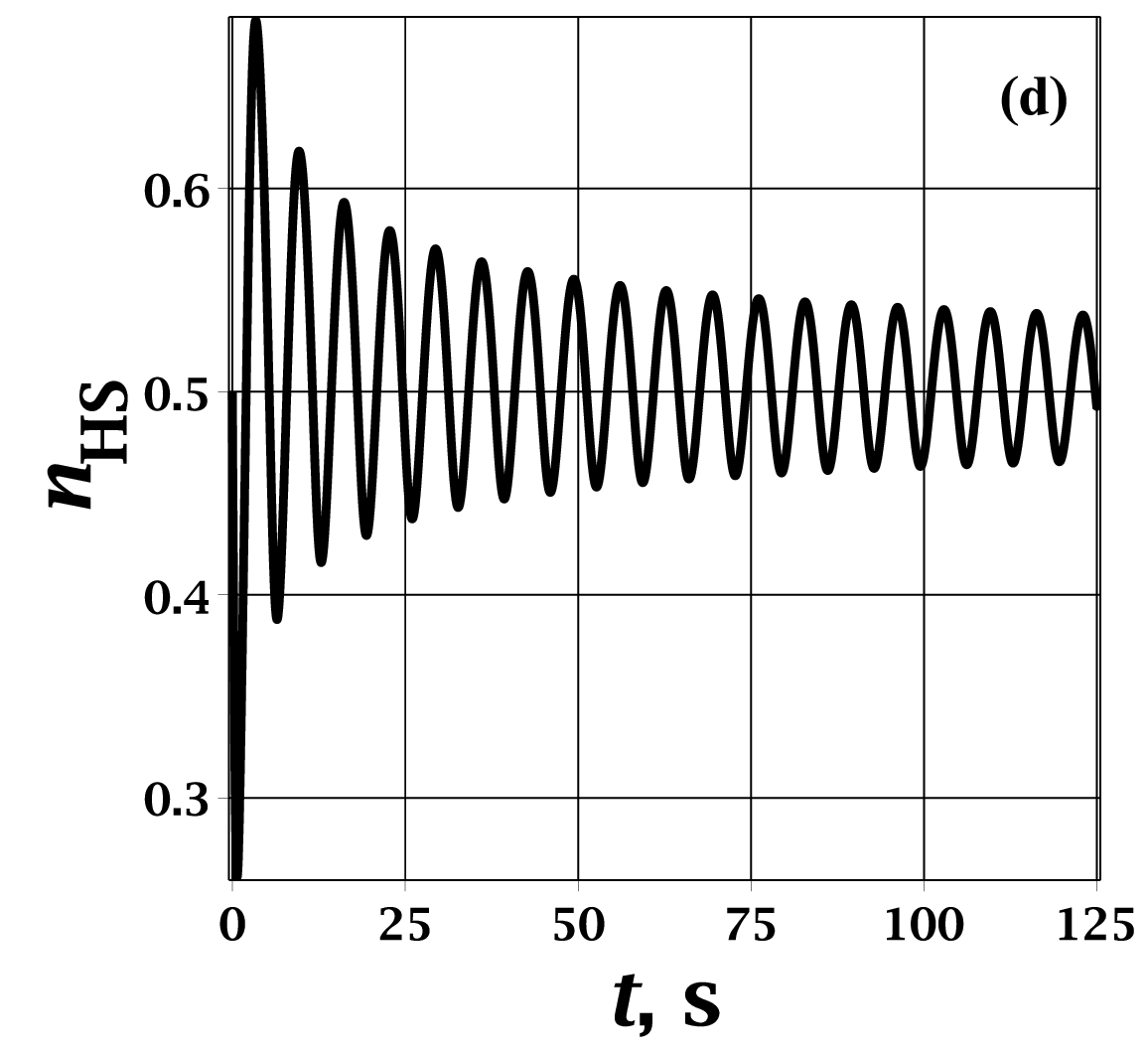}
\caption{\label{fig6} $\alpha = 2.8$~s$^{-1}$,  center. The global phase portrait (Fig.~(a)) shows a spiraling-in spiral. The approach of the spiral to the closed ellipse shown by the red line in Fig.~(b) on an enlarged scale becomes asymptotic (Figs.~(c) and~(d)), for which the approach time tends to infinity.}
\end{figure*}

\begin{figure*}
\centering
\includegraphics[width=0.30\textwidth]{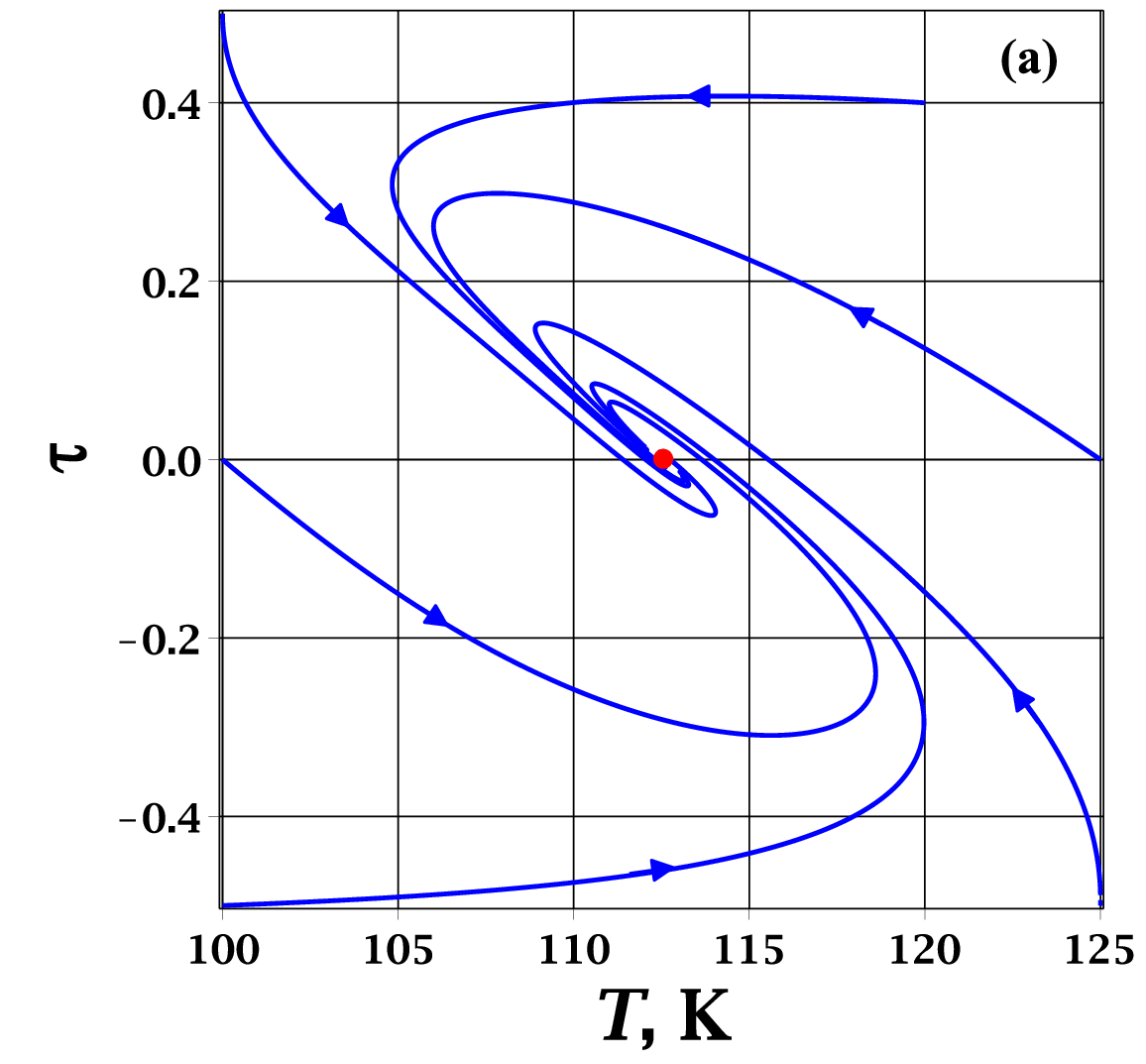}
\hfill 
\includegraphics[width=0.30\textwidth]{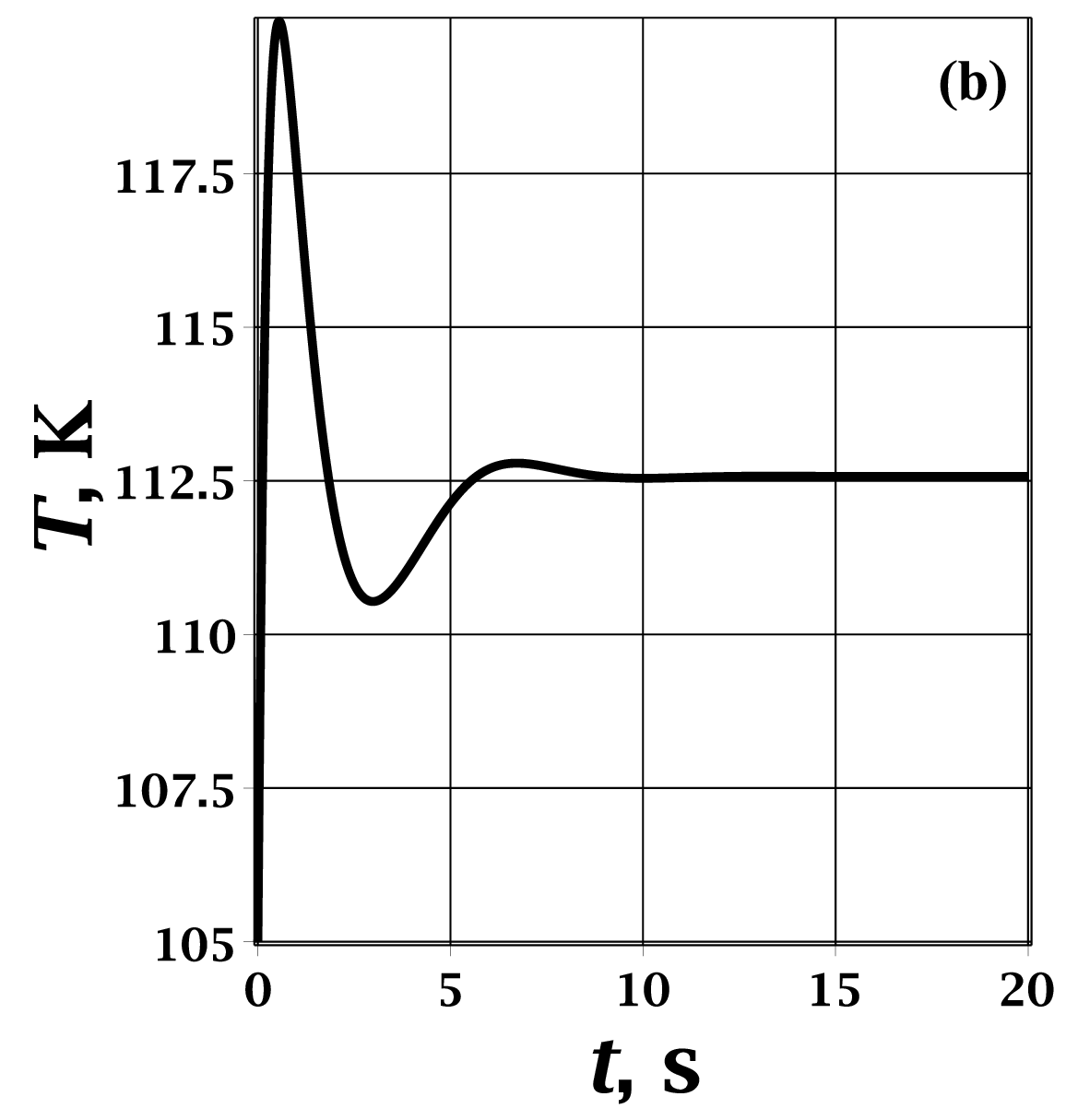}
\hfill 
\includegraphics[width=0.30\textwidth]{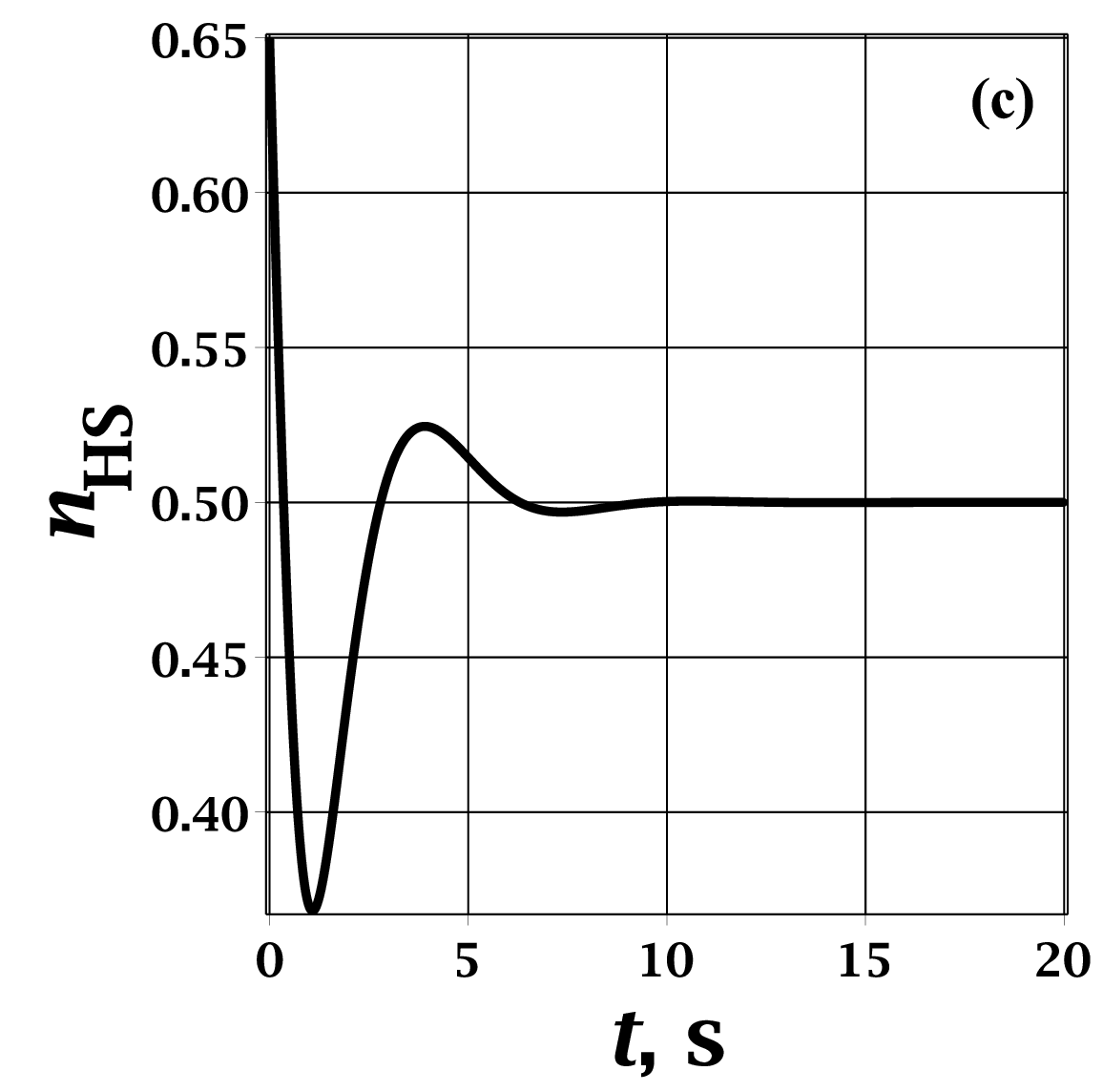}
\caption{\label{fig7} $\alpha = 4$~s$^{-1}$, stable focus. (a) Phase portrait. There is no limit cycle. The characteristic dependences of $T(t)$ and ${n_{HS}}(t)$ are shown in Figs.~(b) and~(c).}
\end{figure*}

\begin{figure*}
\centering
\includegraphics[width=0.30\textwidth]{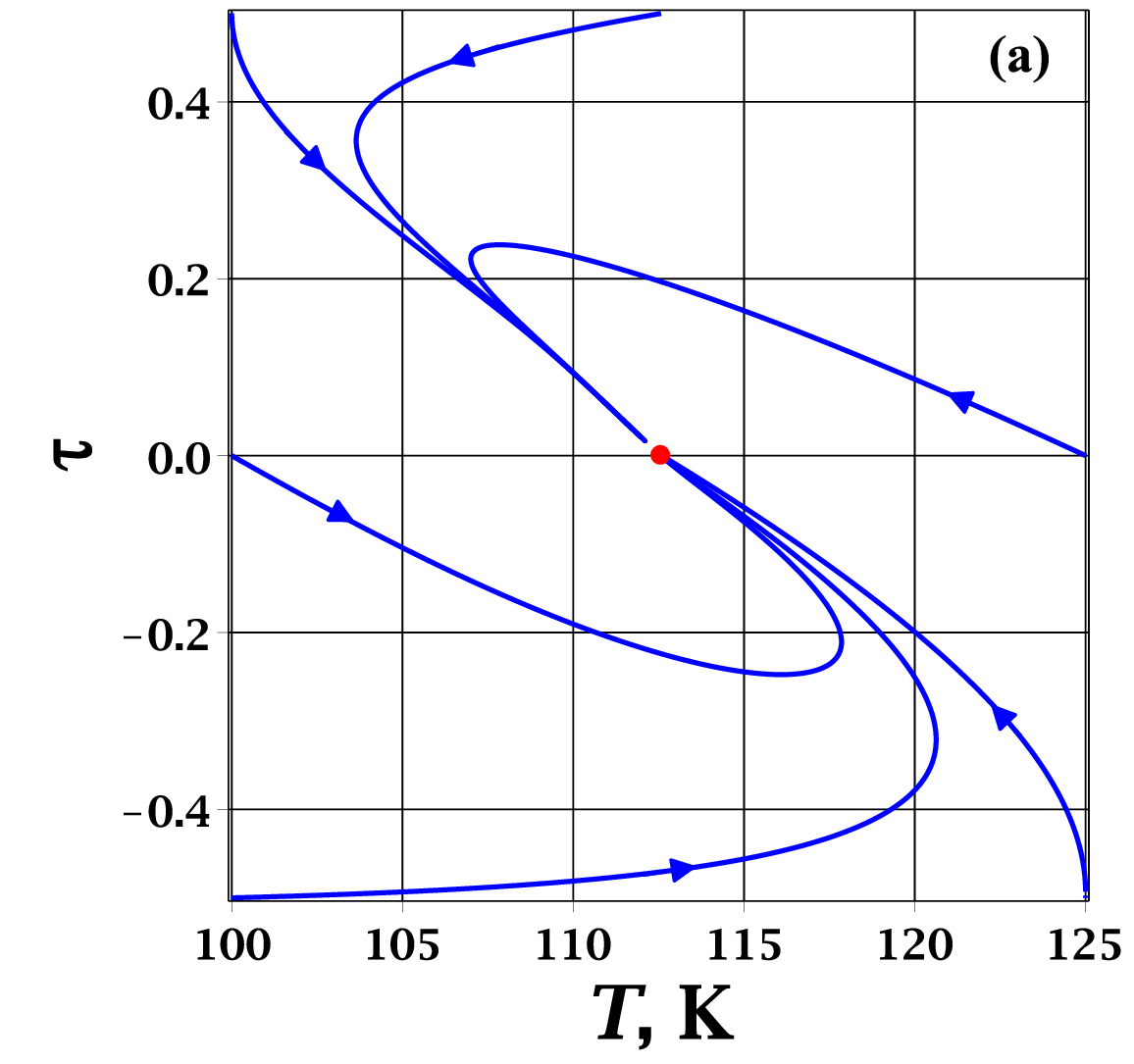}
\hfill 
\includegraphics[width=0.30\textwidth]{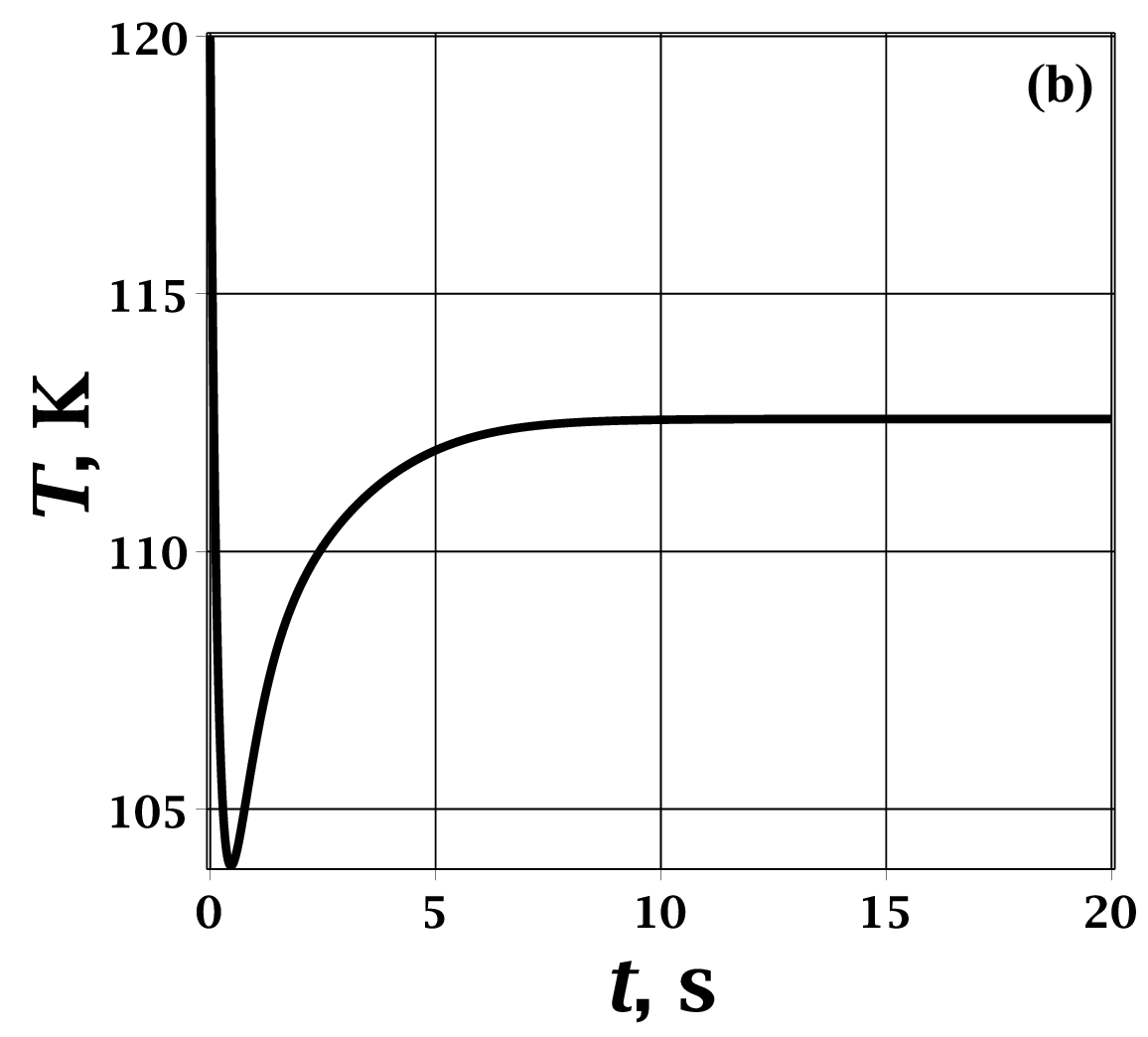}
\hfill 
\includegraphics[width=0.30\textwidth]{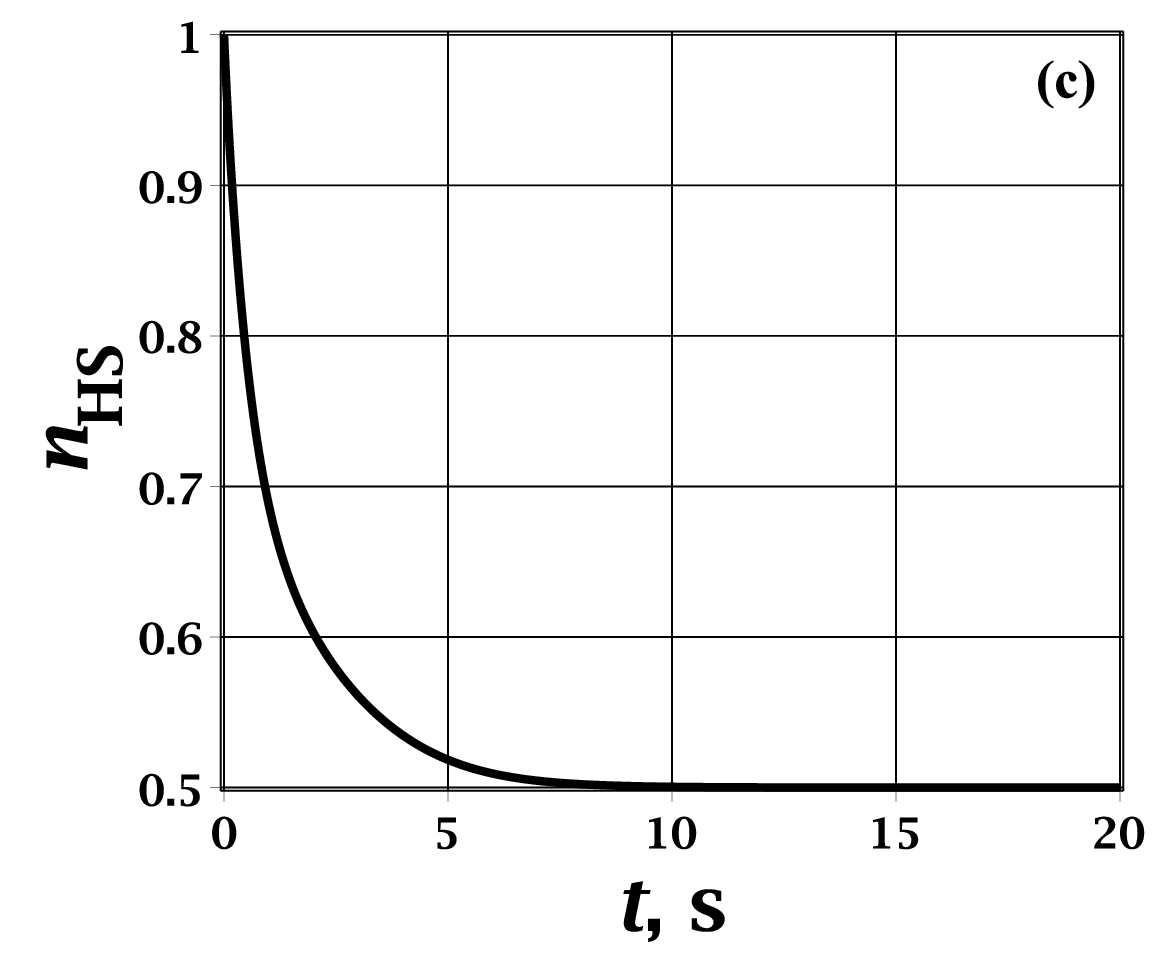}
\caption{\label{fig8} $\alpha = 6$~s$^{-1}$, stable node. (a) Phase portrait. There is no limit cycle. The characteristic dependences of $T(t)$ and ${n_{HS}}(t)$ are shown in Figs.~(b) and~(c).}
\end{figure*}

We seek the solution of the system of linear differential equations~\eqref{Lin_eq} in the form
$\delta \tau \left( t \right) = {C_1}{e^{{\lambda _ + }t}} + {C_2}{e^{{\lambda _ - }t}}$, $\delta T\left( t \right) = {C_3}{e^{{\lambda _ + }t}} + {C_4}{e^{{\lambda _ - }t}}$,
where ${\lambda _ \pm }$ are the eigenvalues of $\Lambda $ satisfying the characteristic equation $Det\left| {\Lambda  - \lambda I} \right| = 0$.
$$
{\lambda _ \pm } = \frac{{Tr\left( \Lambda  \right) \pm \sqrt {{D_\lambda }\left( \Lambda  \right)} }}{2},
$$
where ${D_\lambda }\left( \Lambda  \right) = T{r^2}\left( \Lambda  \right) - 4Det\left( \Lambda  \right)$ is the discriminant of the characteristic equation, $Tr\left( \Lambda  \right) = \Lambda_{11} + \Lambda_{22}$ is the trace of $\Lambda$, and $Det\left( \Lambda  \right) = \Lambda_{11} \Lambda_{22} - \Lambda_{21} \Lambda_{12}$ is its determinant. Using~\eqref{Lambda_ij_1}, we obtain:
\begin{subequations} \label{Tr_Det_Js0_K0}
\begin{eqnarray}
Tr\left( \Lambda  \right) = \Gamma {J_\tau }\left( {{\beta _{eq}}\frac{1}{4}{J_\tau } - 1} \right) - \alpha, \\
Det\left( \Lambda  \right) = \alpha \Gamma {J_\tau }\frac{{\gamma \ln g}}{{2{T_{eq}}}}\left( {T_R^C - {T_R}} \right),
\end{eqnarray}
\end{subequations}
where $T_R^C = {T_{eq}}\left( {1 - 2\frac{{\frac{1}{4}{\beta _{eq}}{J_\tau } - 1}}{{\gamma \ln g}}} \right)$ is the critical value of the thermostat temperature. For $T \ge T_R^C$, the determinant  $Det\left( \Lambda  \right) \le 0$, so the existence of solutions in the form of limit cycles becomes impossible (the condition $Det\left(\Lambda\right) < 0$ corresponds to unstable saddles). For the [Fe(NCSe)(py)$_{22}$(m-bpypz)] single crystal, we have $T_R^C \approx 78.8$~K, $\Gamma = 1/19$~K$^{-1}$~s$^{-1}$, $J_\tau = 152$~K, $\ln{g} = 7$, $\rho = 0.5$, $T_{eq} = 112.6$~K, $\Delta_S = 788$~K~\cite{PhysRevB.89.224303.2014}. Let ${T_R} = 75$~K~\cite{PhysRevB.89.224303.2014}. We plot all stationary points on the bifurcation diagram (Fig.~\ref{fig3}), where the parabola $Det\left( \Lambda  \right) = \frac{{T{r^2}\left( \Lambda  \right)}}{4}$, corresponding to the condition of multiple roots ${D_\lambda }\left( \Lambda  \right) = 0$, shows the boundary between nodes and foci. The regions of nodes and foci are labeled accordingly. The vertical blue line $Tr\left( \Lambda  \right) = 0$ (the ordinate axis) is the position of centers. The regions of stability and instability are highlighted in red and blue, respectively. The stationary points are shown as a straight line, whose equation in the coordinates $Tr \left ( \Lambda\right )$~versus ~$Det \left ( \Lambda\right )$ can be obtained by eliminating $\alpha$ from the two equations~\eqref{Tr_Det_Js0_K0}. The types of stationary points depend on the value of the parameter $\alpha $. The blue boundary point  $\alpha  = 2.8$~s$^{-1}$ is a center; the black boundary points are degenerate nodes.  $Det\left(\Lambda\right) = 0$ at $\alpha = 0$. The small open circles correspond to integer values $\alpha = 1,2,4$ and 6~s$^{-1}$, for which Figs.~\ref{fig4}~--~\ref{fig8} show the $T - \tau $ phase portraits of Eqs.~\eqref{main_dT/dt} and~\eqref{main_dtau/dt_Js0} with $\frac{{\Delta H}}{{{C_P}}} = 0$  and the corresponding time dependences $T\left( t \right)$ and $n_{HS} \left( t \right)$ for different initial conditions.

\begin{figure}
\centering
\includegraphics[width=9.0cm]{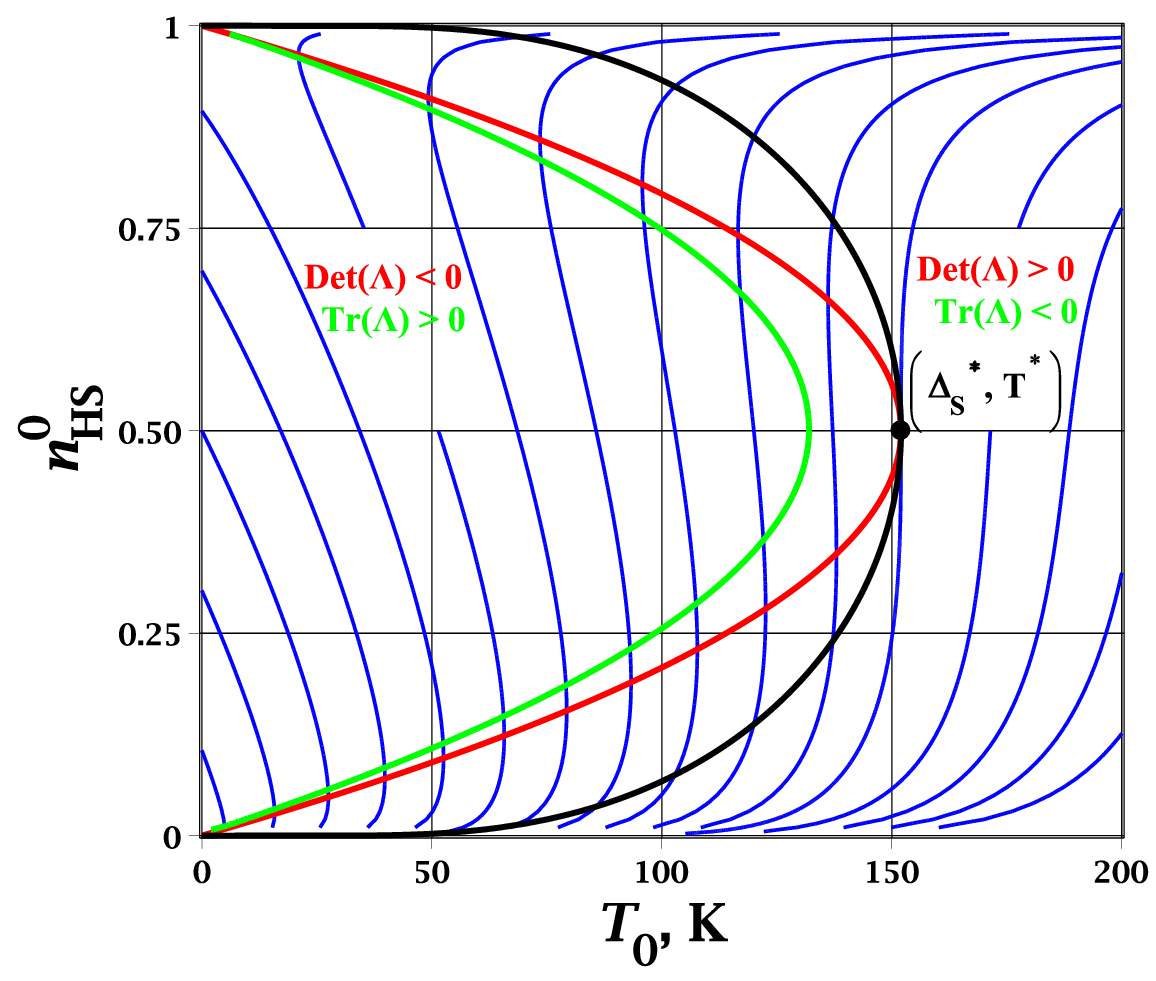}
\caption{\label{fig9} Temperature dependence of the HS state population ${n^0_{HS}} = {\tau _0} + \frac{1}{2}$ for different values of the spin gap ${\Delta _S}$: from $-240$ (leftmost curve) to 1560~K  (rightmost curve). The black line is the binodal, the boundary of the first-order phase transition region. The red line ($Det\left( \Lambda \right) = 0$) is the spinodal, it separates the region of metastable states from absolutely unstable (labile) states. The green line is $Tr\left( \Lambda \right) = 0$ for $\alpha = 1$~s$^{-1}$ and $\frac{{\Delta H}}{{{C_P}}} = 1$~K.  $\left( {\Delta _S^ * ,{T^*}} \right)$ is the critical point (see Fig.~\ref{fig1}b): $\Delta_S^* = 1064$~K, $T^* = 152$~K.}
\end{figure}

The existence of the limit cycles shown in Figs.~\ref{fig4} and~\ref{fig5} turned out to be possible due to the external radiation $I$, which provides feedback in Eqs.~\eqref{main_dT/dt} and \eqref{main_dtau/dt_Js0}. If, however, we consider the case where $I = 0$ but $\frac{{\Delta H}}{{{C_P}}} \ne 0$ (which is interesting because then feedback is possible without external action and arises due to nonequilibrium processes within the system itself), then for ${J_S} = 0$ the existence of limit cycles becomes impossible. This can be easily verified by examining Fig.~\ref{fig9}, where the blue lines show a set of equilibrium solutions of the self-consistent equation~\eqref{stationary_tau0_JS0} for different values of the spin gap ${\Delta _S}$: from  $-240$ (leftmost curve) to 1560~K (rightmost curve). The black solid line in Fig.~\ref{fig9} shows the boundary of the first-order phase transition region -- the binodal (see Fig.~\ref{fig1}). The red line, the spinodal, separates the region of metastable states from absolutely unstable (labile) states. The equation of the spinodal can be obtained from the condition $Det\left( \Lambda  \right) = 0$, if in the Jacobian matrix  $\Lambda$ we set  $g\left( {\tau ,T} \right) = - \alpha \left( {T - {T_R}} \right) - \frac{{\Delta H}}{{{C_p}}}\frac{{\partial \tau }}{{\partial t}}$, and consider the derivatives of $f$ and $g$ at the point $\left( {{T_0}, {\tau _0}} \right)$. Using~\eqref{stationary_tau0_JS0}, we obtain:
\begin{eqnarray}
\begin{split}
&{\Lambda _{11}} = {\left. {\frac{{\partial f}}{{\partial \tau }}} \right|_0} = \Gamma {J_\tau }\left[ {\left( {1 - 4\tau _0^2} \right){\beta _0}\frac{{{J_\tau }}}{4} - 1} \right], \\
&{\Lambda _{12}} = {\left. {\frac{{\partial f}}{{\partial T}}} \right|_0} = \frac{1}{{{T_0}}}\left( {{\Lambda _{11}} + \Gamma {J_\tau }} \right)\left( {\frac{{{\Delta _S}}}{{{J_\tau }}} - {\tau _0}} \right), \\
&{\Lambda _{21}} = {\left. {\frac{{\partial g}}{{\partial \tau }}} \right|_0} =  - \frac{{\Delta H}}{{{C_P}}}{\left. {\frac{{\partial f}}{{\partial \tau }}} \right|_0} = - \frac{{\Delta H}}{{{C_P}}}{\Lambda _{11}}, \\
&{\Lambda _{22}} = {\left. {\frac{{\partial g}}{{\partial T}}} \right|_0} =  - \alpha  - \frac{{\Delta H}}{{{C_P}}}{\left. {\frac{{\partial f}}{{\partial T}}} \right|_0} =  - \alpha  - \frac{{\Delta H}}{{{C_P}}}{\Lambda _{12}}.
\end{split}
\label{Lambda_ij_2}
\end{eqnarray}
Equation \eqref{Lambda_ij_2} readily yields
$$
Tr\left( \Lambda  \right) = {\Lambda _{11}} - \frac{{\Delta H}}{{{C_p}{T_0}}}\left( {{\Lambda _{11}} + \Gamma {J_\tau }} \right)\left( {\frac{{{\Delta _S}}}{{{J_\tau }}} - {\tau _0}} \right) - \alpha, 
$$
$$
Det\left( \Lambda  \right) = - \alpha {\Lambda _{11}}.
$$
If $Det\left( \Lambda  \right) = 0$, then either $\Lambda _{11} = 0$ or $\tau _0^2 = \frac{1}{4} - \frac{{{k_B}{T_0}}}{{{J_\tau }}}$ -- the equation of the spinodal. For absolutely unstable solutions, $Det\left( \Lambda  \right) < 0$. Everywhere outside the region bounded by the spinodal, $Det\left( \Lambda  \right) > 0$. For $\alpha  = 0$ and $\frac{{\Delta H}}{{{C_P}}} = 0$, the condition $Tr\left( \Lambda  \right) = 0$ holds only for $T_0$ and $\tau_0$ belonging to the spinodal, and the region $Tr\left( \Lambda  \right)~>~0$ ($Tr\left( \Lambda  \right) < 0$) coincides with the region $Det\left( \Lambda  \right) < 0$ ($Det\left( \Lambda  \right) > 0$). For any nonzero $\alpha $ and/or $\frac{{\Delta H}}{{{C_P}}}$ (only positive values have physical meaning), the boundary $Tr\left( \Lambda  \right) = 0$, shown for example by the green line in Fig.~\ref{fig9} for $\alpha = 1$~C$^{-1}$ and $\frac{{\Delta H}}{{{C_P}}} = 1$~K, always lies inside the region $Det\left( \Lambda  \right) < 0$, so that the region $Tr\left( \Lambda  \right) > 0$ always lies inside the region $Det\left( \Lambda  \right) < 0$. As $\alpha $ or/and $\frac{{\Delta H}}{{{C_P}}}$ increase, the region $Tr\left( \Lambda  \right) > 0$  shrinks. For $\alpha  = 0$ and $\frac{{\Delta H}}{{{C_P}}} = 0$, the green line coincides with the red line. It is seen from Fig.~\ref{fig9},  that the necessary conditions  $Det\left( \Lambda  \right) > 0$ and $Tr\left( \Lambda  \right) > 0$ for the existence of unstable solutions and limit cycles are not satisfied when $I = 0$.

Completely opposite behavior can be observed if we set $J_\tau$ = 0, but ${J_S} \ne 0$. In this case, numerical solution of Eqs.~\eqref{main_dT/dt} and~\eqref{main_(dtau/dt)_(dm/dt)} for $\frac{{\partial m}}{{\partial t}} = \frac{{\partial \tau }}{{\partial t}} = \frac{{\partial T}}{\partial t} = 0$ shows that, for the entire range of parameters ${\Delta _S}$ and ${T_0}$, where ${m_0} \ne 0$, one can assume with good accuracy that $\tau_0 = \frac{1}{2}$ or ${n^0_{HS}} = {\tau _0} + \frac{1}{2} = 1$ (Fig.~\ref{fig2}). Therefore, in the stationary case and for ${m_0} \ne 0$, the presence of photothermal heating $I$ essentially reduces to a redefinition of the thermostat temperature ${T_R}$ (see~\eqref{main_dT/dt}.
Taking this into account, from Eq.~\eqref{main_dm/dt} we obtain:
\begin{equation}
    {m_0} = \frac{{{\theta _0}}}{{{e^{\beta {\Delta _{\tau 0}}}} + {\theta _0}}}S{B_S}\left( {{\beta _0}S{J_S}{m_0}} \right),
    \label{m0_nHS=1}
\end{equation}
where ${\Delta _{\tau 0}} = {\Delta _S} - {k_B}T\ln {g_\tau }$,
$$
{\theta _0} = \frac{{\sinh \left[ {\left( {2S + 1} \right){\beta _0}{J_S}{{{m_0}} \mathord{\left/
 {\vphantom {{{m_0}} 2}} \right.
 \kern-\nulldelimiterspace} 2}} \right]}}{{\sinh \left[ {{\beta _0}{J_S}{{{m_0}} \mathord{\left/
 {\vphantom {{{m_0}} 2}} \right.
 \kern-\nulldelimiterspace} 2}} \right]}}.
$$

\begin{figure*}
\centering
\includegraphics[width=18.0cm]{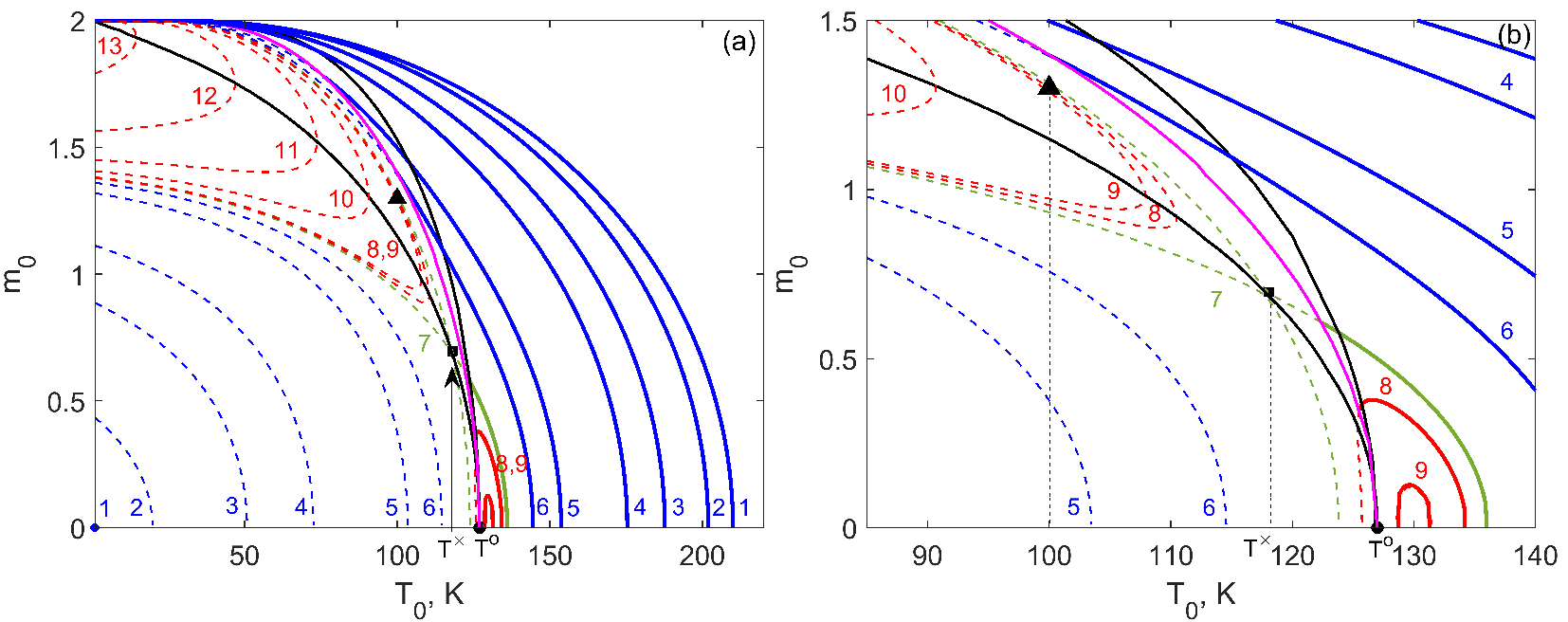}
\caption{\label{fig10} (a) Temperature dependence of the solutions of Eq.~\eqref{m0_nHS=1} for different values of the spin gap ${\Delta _S}$. (b) Enlarged view of panel (a) in the vicinity of the special points. The solutions for $\Delta_S = 0$, 100, 200, 250, 296, and 305~K are indicated in blue and marked by the integers 1 to 6, respectively. The solutions for $\Delta_S = 309.5$, 309.85, 315, 325, 350, and 400~K are indicated in red and marked by the integers 8 to 13, respectively. The solutions for $\Delta_S = \Delta_S^\times = 309.08$~K (at $\Delta = \Delta^\times$) are highlighted in green and marked by the number 7. The solid lines show the solutions corresponding to the global minimum of the free energy $F$. The black solid lines indicate the boundaries of the first-order phase transition region — the binodal. The lower (upper) black curve corresponds to the right (left) boundary in Fig.~\ref{fig2}. The pink solid line is the first-order phase transition line. The symbols \scalebox{0.9}{$\blacksquare$} and \scalebox{1.5}{$\bullet$} correspond to the positions of the points $\left( {\Delta^ \times, {T^ \times }} \right)$ and $\left( {\Delta^ \odot, {T^ \odot }} \right)$ in Fig.~\ref{fig2}, $\blacktriangle$ denotes the stationary point ($T_0 = 100$~K, $m_0 = 1.3$) in the vicinity of which a limit cycle was found (Fig.~\ref{fig11}).}
\end{figure*}

Figure~\ref{fig10} shows the temperature dependence of the solutions of Eq.~\eqref{m0_nHS=1} for different values of the spin gap ${\Delta _S}$. The solutions for $\Delta_S = 0$, 100, 200, 250, 296 and 305~K are indicated in blue and marked by the integers 1 to 6, respectively. The solutions for $\Delta_S = 309.5$, 309.85, 315, 325, 350, and 400~K are indicated in red and marked by the integers 8 to 13, respectively. The solutions for $\Delta_S = \Delta_S^\times = 309.08$~K (at $\Delta = \Delta^\times$) are highlighted in green and marked by the number 7. The solid lines show the solutions corresponding to the global minimum of the free energy $F$. The black solid lines indicate the boundaries of the first-order phase transition region (see Fig.~\ref{fig2}), the binodal. The lower (upper) black curve corresponds to the right (left) boundary in Fig.~\ref{fig2}. The pink solid line is the first-order phase transition line. The symbols \scalebox{0.9}{$\blacksquare$} and \scalebox{1.5}{$\bullet$} correspond to the positions of the points $\left( {\Delta^ \times, {T^ \times }} \right)$ and $\left( {\Delta^ \odot, {T^ \odot }} \right)$ in Fig.~\ref{fig2}, respectively.

Numerical calculations show that the lower black curve in Fig.~\ref{fig10} coincides with the position of the spinodal and is determined by the condition $Det\left( \Lambda  \right) = 0$ for the system of Eqs.~\eqref{main_dT/dt} and \eqref{main_(dtau/dt)_(dm/dt)} with $I = 0$. Inside the region bounded by the spinodal (below the lower black curve), $Det\left( \Lambda  \right) < 0$. Everywhere outside this region, $Det\left( \Lambda  \right) > 0$. Therefore, the existence of limit cycles is possible in the vicinity of stationary points located in the first-order phase transition region bounded by the two black curves (Fig.~\ref{fig10}).

\begin{figure*}
\centering
\includegraphics[width=0.30\textwidth]{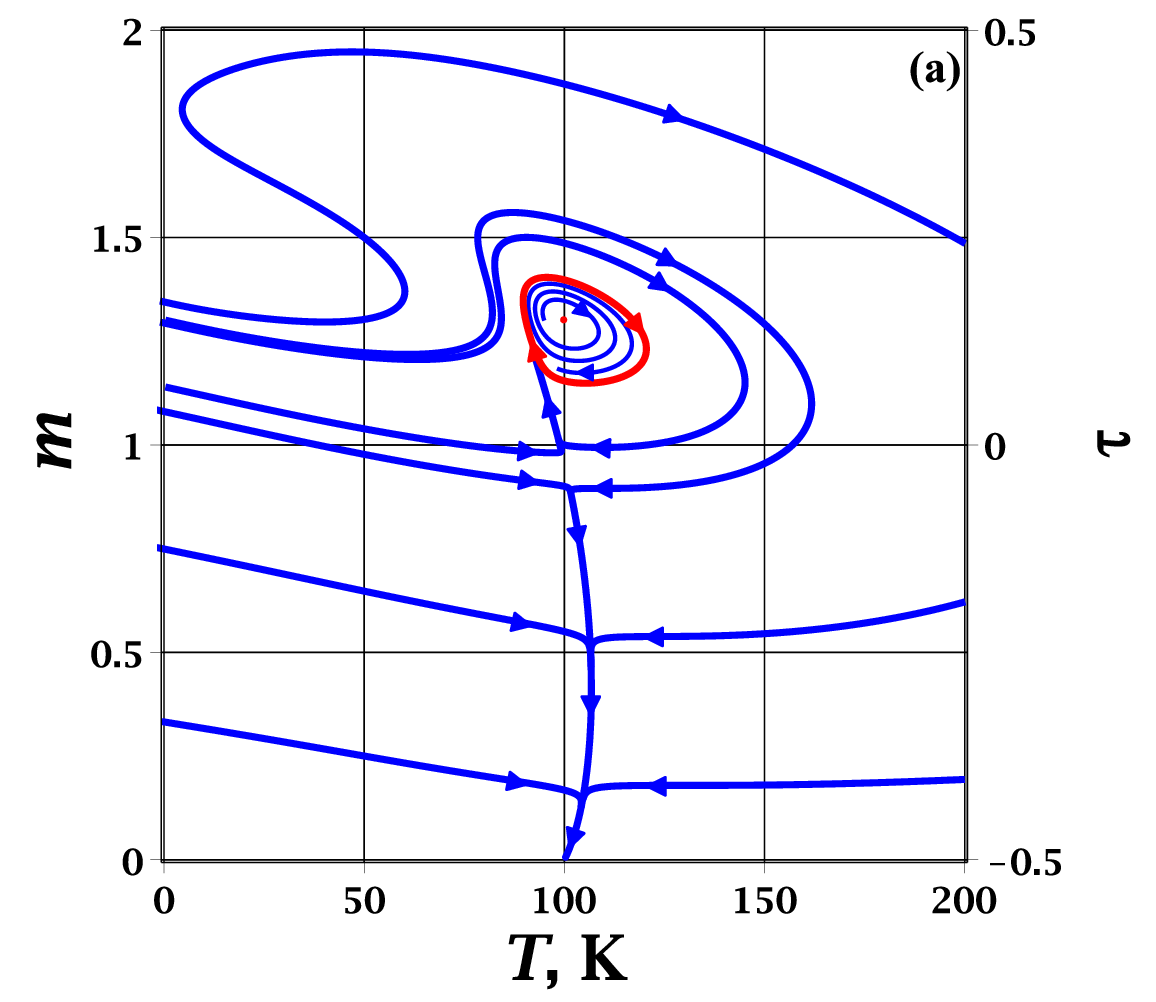}
\hfill 
\includegraphics[width=0.30\textwidth]{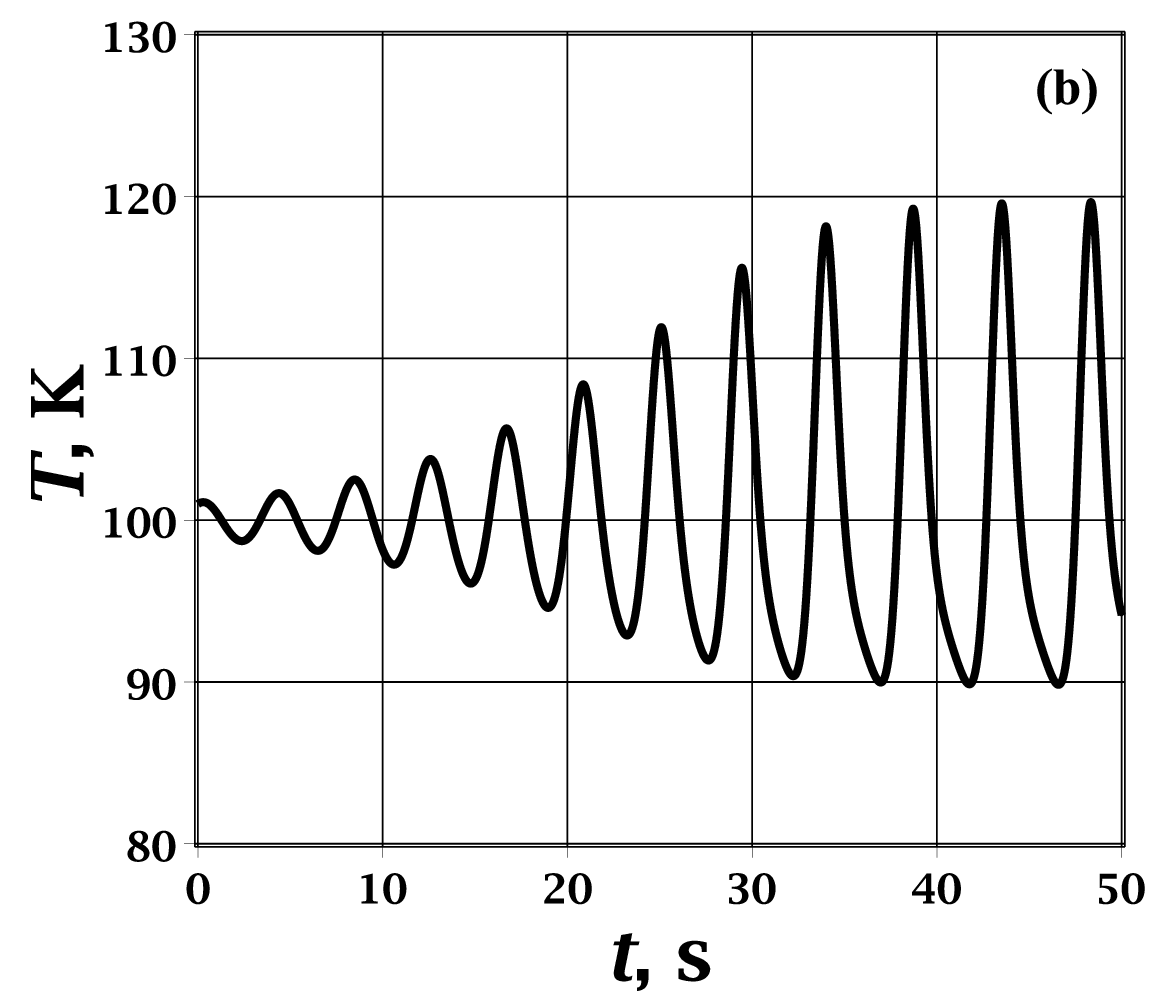}
\hfill 
\includegraphics[width=0.30\textwidth]{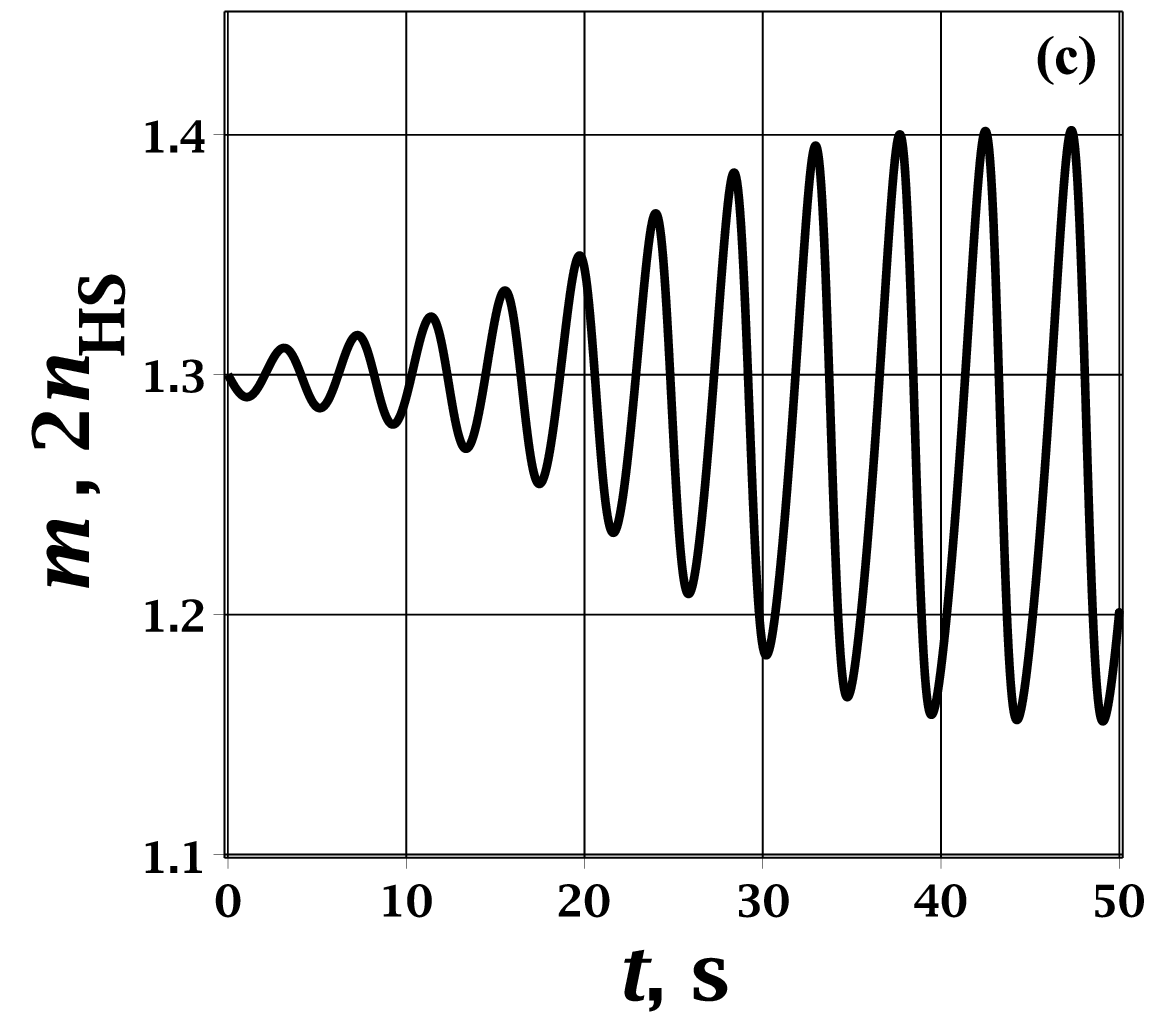}
\\ 
\hfill 
\hfill 
\hfill 
\hfill 
\hfill 
\hfill 
\hfill 
\hfill 
\includegraphics[width=0.30\textwidth]{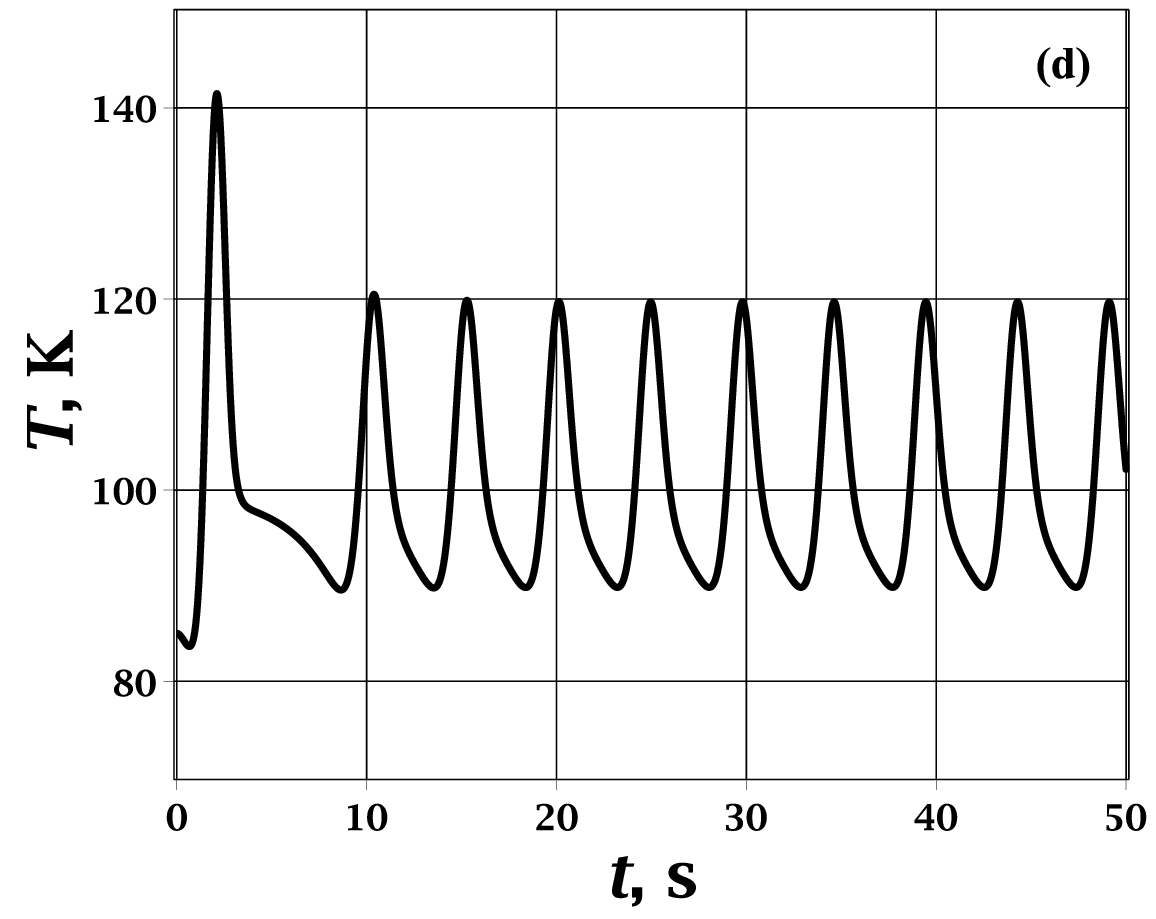}
\hfill 
\includegraphics[width=0.30\textwidth]{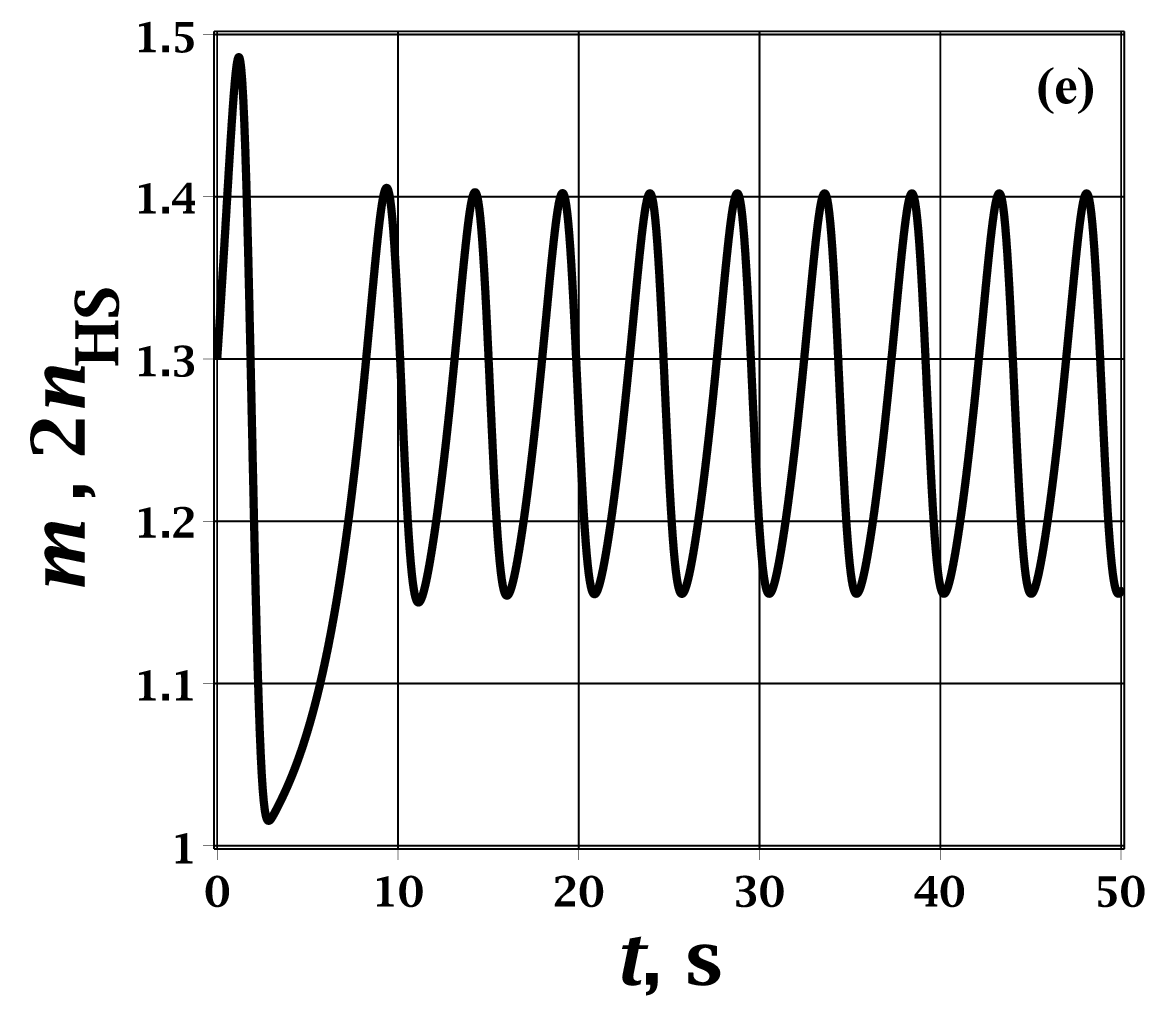}
\caption{\label{fig11} (a) Phase portrait of the system of differential equations \eqref{main_dT/dt} and \eqref{main_(dtau/dt)_(dm/dt)} for the case $J_\tau = 0$ and $I = 0$ in the vicinity of the stationary point ($T_0 = 100$~K, $m_0 = 1.3$) corresponding to the spin gap $\Delta_S = 309.5$~K. The red line shows the limit cycle. Self-oscillations of temperature $T$, magnetization $m$, and HS state population $n_{HS}$ for initial conditions chosen inside and outside the limit cycle are shown in (b,c) and (d,e), respectively. The calculations were performed for the following set of parameters: $\Gamma = 1/20$~K$^{-1}$~C$^{-1}$, $\alpha = 4.5$~C$^{-1}$ and $\frac{{\Delta H}}{{{C_P}}} = 100$~K.}
\end{figure*}

As an example, Fig.~\ref{fig11} shows the results of numerical solution of Eqs.~\eqref{main_dT/dt} and \eqref{main_(dtau/dt)_(dm/dt)} for $I = 0$ in the vicinity of the stationary point (${T_0} = 100$~K,  ${m_0} = 1.3$), marked by the triangle $\blacktriangle$ in Fig.~\ref{fig10}. Self-sustained oscillations of temperature (Fig.~\ref{fig11}b,d) and magnetization, generating oscillations of the population of the $^5T_{2g}$ multielectron term (Fig.~\ref{fig11}c,e), are clearly visible. Switching on the external radiation $I$ leads to a decrease in the amplitude of these oscillations.

\section{Discussion and conclusions}
The principal difference between the two cases considered, $J_S = 0$ ($J_\tau \ne 0$) and $J_\tau = 0$ ($J_S \ne 0$), is that in the former, self-oscillations of the HS state population $n_{HS}$ and temperature $T$ are impossible without external stationary radiation providing photothermal heating and feedback. In the latter case, self-sustained oscillations of the magnetization $m$, population $n_{HS}$, and temperature $T$ are possible without an external energy source and occur via a mechanism similar to autocatalytic reactions. Despite the differences between Eqs.~\eqref{main_dT/dt} and \eqref{main_(dtau/dt)_(dm/dt)} and the Belousov--Zhabotinsky equations, magnetically ordered SC systems exhibit nonlinear mechanisms that ensure periodic transitions from one state to another (self-oscillations).

Since the LS state is nonmagnetic, it is evident that a change in the population $n_{HS}$ will inevitably affect the magnetization $m$ of the material. We have succeeded in demonstrating a much less obvious inverse dynamical effect, namely, that a change in the magnetization $m$ leads to a change in the population $n_{HS}$ in the presence of only the interatomic exchange interaction $J_S$ acting in the spin subspace of the states. In other words, changes in the spin subsystem induce a response in the orbital subsystem.

The results obtained in this work for magnetically ordered SC compounds with $3d^6$ electron configuration of transition metal ions do not lose their generality and can be readily applied to any magnetically ordered SC systems. In our opinion, the most suitable example of a compound in which experimental observation of magnetization self-oscillations is possible is the FeBO$_3$ single crystal. Iron borate single crystals are well studied, have a rather high Néel temperature, and are transparent in the visible spectral range. It is known that they lose their transparency upon the pressure-induced transition from the HS to LS state~\cite{Lyubutin_UFN_2009}.

\textbf{Acknowledgments}
This work was performed within the state assignment of the Kirensky Institute of Physics of the Siberian Branch of the Russian Academy of Sciences.

\bibliography{manuscriptbibl_Autocatalytic2}
\end{document}